\documentclass{IEEEtaes_arxiv}

\usepackage{color,array,amsthm}
\usepackage{graphicx}
\usepackage{tabularx} % Table
\usepackage{booktabs}   % 添加这行
\usepackage{amsmath,amsfonts}
\usepackage{algorithmic}
\usepackage{algorithm}

\usepackage{array}
\usepackage{textcomp}
\usepackage{stfloats}
\usepackage{url}
\usepackage{verbatim}
\usepackage{graphicx}
\usepackage{cite}
\usepackage{multirow} % for cmd 'multirow', 'multicolumn'
\usepackage{subcaption}
\usepackage{algorithm}      % 算法环境
\usepackage{algorithmic}    % 算法排版（注意：这是algorithmic，不是algpseudocode）
\usepackage{amsmath}        % 数学符号

\usepackage{float}

\begin{document}
% \nocite{*}
\title{Robust DOA Estimation for GNSS-like Signals Under Jamming: A Rotation-Enhanced SWN-MUSIC Approach}

\author{Jiyao Jiang}
\author{Xiaowei Cui}
\author{Gang Liu}
\author{Sihao Zhao}
\author{Mingquan Lu}
\affil{Department of Electronic Engineering, Tsinghua University, Beijing, China
\\ 
State Key Laboratory of Space Network and Communications, Tsinghua University, Beijing, China
\\
National Research Center for Information Science and Technology, Beijing, China
}

%% \author{FOURTH D. AUTHOR}
%% \affil{University of Colorado, Colorado, USA}

\receiveddate{This work has been submitted to the IEEE for possible publication. Copyright may be transferred without notice, after which this version may no longer be accessible.\\
This work was supported by the National Natural Science Foundation of China (NSFC) under Grant 62541105.}
%% \accepteddate{XXXXX XX XXXX}
%% \publisheddate{XXXXX XX XXXX}

\corresp{{\itshape (Corresponding author: Xiaowei Cui)} }

\authoraddress{Jiyao Jiang and Gang Liu are with the Department of Electronic Engineering, Tsinghua University, Beijing 100084, China (e-mail: jiang-jy23@mails.tsinghua.edu.cn; liu\_gang@tsinghua.edu.cn). Xiaowei Cui and Mingquan Lu are with the Department of Electronic Engineering, Tsinghua University, Beijing 100084, China, the State Key Laboratory of Space Network and Communications, Tsinghua University, Beijing 100084, China, and the National Research Center for Information Science and Technology, Beijing 100084, China (e-mail: cxw2005@tsinghua.edu.cn; lumq@tsinghua.edu.cn). Sihao Zhao is an independent researcher in Alberta, Canada (e-mail: zsh01@tsinghua.org.cn).}

% \thanks{Chenxin Tu and Gang Liu are with the Department of Electronic Engineering, Tsinghua
% University, Beijing 100084, China (e-mail: tcx22@mails.tsinghua.edu.cn; liu\_gang@tsinghua.edu.cn).}
% \thanks{Xiaowei Cui and Mingquan Lu are with the Department of Electronic Engineering, State Key Laboratory of Space Network and Communication, Tsinghua University, Beijing 100084, China (e-mail:
% cxw2005@tsinghua.edu.cn; lumq@tsinghua.edu.cn).}

\editor{}
\supplementary{}

\markboth{JIANG ET AL.}{Robust DOA Estimation for GNSS Signals Under Jamming: A Rotation-Enhanced SWN-MUSIC Approach}
\maketitle
\begin{abstract}
In jamming environments, accurate direction-of-arrival (DOA) estimation for Global Navigation Satellite System (GNSS)-like signals using antenna arrays is crucial for spoofing detection and subsequent countermeasures.
However, estimation accuracy strongly depends on the incident direction of the signal of interest (SOI) under a given jamming and array configuration, with severe errors occurring in certain directions---a phenomenon whose physical mechanisms remain unclear.
To address this gap, we identify two types of geometric ill-conditions between the array manifold and the jamming subspace, clarifying that certain directions suffer from inherently unfavorable observation geometry, which cannot be overcome by algorithmic refinement alone.
Motivated by this insight, we introduce a  paradigm namely rotational diversity, which treats the array attitude as a controllable sensing resource to actively reconfigure the observation geometry, thereby providing complementary perspectives in which the problematic directions can be accurately estimated.
To implement this paradigm, we develop a  Rotation-Enhanced Spatial Whitening and Normalized Multiple Signal Classification (SWN-MUSIC) framework, which performs independent DOA estimations across multiple rotation angles and selectively integrates reliable estimates.
Simulation results demonstrate substantial improvement in accuracy in the directions that were poorly estimated under the fixed-array SWN-MUSIC, while maintaining performance in well-conditioned directions, thereby producing more robust DOA estimates.

\end{abstract}

\begin{IEEEkeywords} global navigation satellite system, direction of arrival, multiple signal classification algorithm, jamming, antenna array.
\end{IEEEkeywords}

\section{INTRODUCTION}     
\IEEEPARstart{G}{lobal} Navigation Satellite Systems (GNSSs) provide critical positioning, navigation, and timing (PNT) services on a global scale, underpinning the fundamental operation of modern society \cite{4745647}.
However, owing to their open-access signal structures and  minute received power, GNSS signals are inherently vulnerable to radio frequency (RF) interference \cite{11165996, 7445815}.
The rising prevalence of security-related incidents has already resulted in severe operational repercussions \cite{jones2017spoofing, androjna2021impact, schmidt2016survey, Lo2026Global}, prompting an urgent need for resilient PNT frameworks to mitigate sophisticated electronic attacks \cite{radovs2024recent, 10.3389/fphy.2023.1133316, 7444136}.

Jamming and spoofing stand as the two primary forms of RF interference.
While defense mechanisms against jamming have become relatively mature \cite{7444116, broumandan2016overview, 7444122}, spoofing presents a more formidable challenge due to its stealth.
Complementary to  non-spatial anti-spoofing approaches such as signal quality monitoring \cite{zhou2022new, chen2022gnss} and power monitoring\cite{akos2012s, deng2022gnss}, direction-of-arrival (DOA)-based spoofing detection \cite{meurer2016direction, 10568334, Danelli2026, marcos2018interference, meurer2012robust, s19183870} has been proven highly effective,
as the DOAs of the authentic satellite signals are extremely difficult to replicate.
Spoofing signals typically exhibit discernible spatial anomalies, such as multiple GNSS signals sharing a common DOA \cite{s19183870}, or the estimated DOAs significantly deviating from ephemeris-based expectations \cite{meurer2012robust}.

The DOA information of GNSS-like signals, including both the authentic and spoofing signals, is valuable not only for spoofing detection but also for enabling subsequent mitigation strategies. 
Specifically, by steering nulls toward the spoofing sources while simultaneously enhancing the authentic signals, beamforming can effectively suppress spoofing signals \cite{daneshmand2012low}.
In addition, DOA information facilitates spoofer localization \cite{rs15235497}, supporting further mitigation of spoofing attacks.

DOA estimation for GNSS-like signals using antenna arrays has attracted extensive research interest. 
In jamming-free environments, the estimation is straightforward, as despreading can sufficiently elevate the signal-to-noise ratio (SNR) to support subsequent conventional subspace-based algorithms \cite{meurer2016direction, appel2015robust, danelli2026gnss}.
However, these methods fail in the presence of jamming because even after despreading, the non-negligible residual jamming remains detrimental to DOA estimation.
Moreover, as jamming and spoofing attacks increasingly occur concurrently \cite{10480364, radovs2024recent, 10719606}, robust DOA estimation in such hostile environments becomes critically important and is the focus of this work.

To address jamming, the widely recognized approach is to perform jamming suppression prior to despreading, followed by subspace-based DOA estimation on the despread signals.
However, applying conventional MUSIC \cite{arribas2023receiver} is inadequate: the suppression process distorts the original signal subspace, causing pseudospectrum peaks to flatten or shift. 
The study in \cite{s19183870} modified the MUSIC pseudospectrum to compensate for these distortions, yet remains prone to generating spurious peaks at jammer directions.
Spatial Whitening and Normalized MUSIC (SWN-MUSIC) \cite{wang2024two} was the first to systematically resolve these issues and establish the Cramér-Rao Lower Bound (CRLB) for this problem. 
By integrating spatial whitening to suppress jamming with a norm-normalization technique, SWN-MUSIC mitigates jamming-induced distortions and achieves performance that surpasses existing methods while approaching the CRLB.

\begin{figure}[htbp]
\centering
\includegraphics[width=0.8\linewidth]{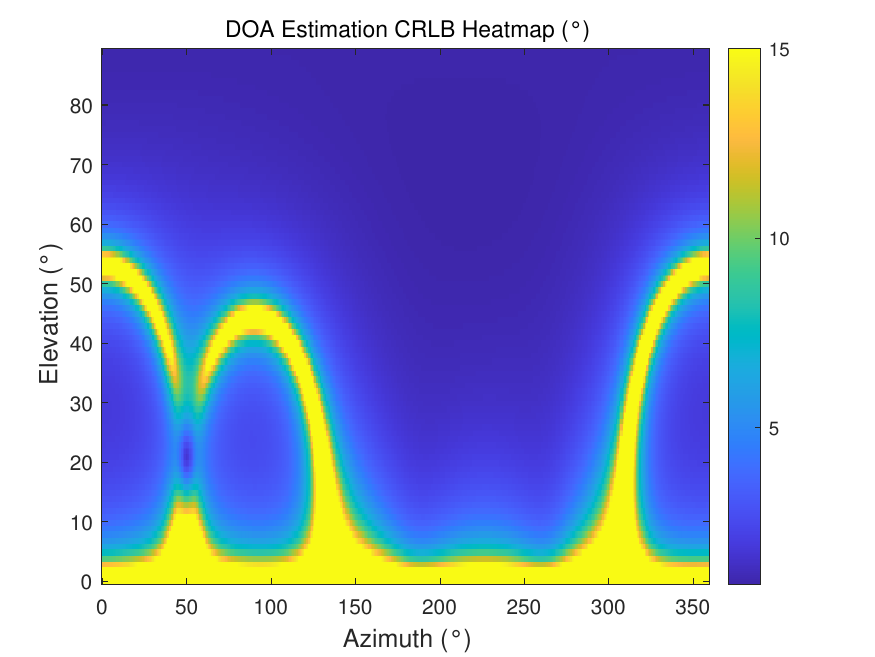}
\caption{Conceptual illustration of  the DIP. Each pixel represents a potential SOI incident direction, with the color indicating the corresponding CRLB.
}
\label{DIP_fig}
\end{figure}

However, as revealed by \cite{11231075}, the CRLB depends heavily on the actual incident direction of the signal of interest (SOI) under a given jamming and array configuration. 
As illustrated in Fig. \ref{DIP_fig}, 
once the SOI originates from the directions corresponding to the high-error bands (highlighted in yellow), the estimation accuracy collapses---a phenomenon defined in this work as the Direction-dependent Inaccuracy Phenomenon (DIP).

To cope with such inaccuracies, \cite{11231075} mitigated the DIP impact in the context of array attitude determination by identifying and down-weighting unreliable DOA estimates.
However, this strategy relies on the measurement redundancy inherent in attitude determination, where a few accurate satellite DOAs are sufficient for array attitude estimation.
Such a passive strategy is unsuitable for spoofing detection, where the SOI DOA itself is the quantity that must be accurately estimated; should an SOI truly originate from inaccurate directions, the resulting erroneous DOA estimates will lead to spoofing detection failure, rendering subsequent countermeasures futile. 
Furthermore, the physical mechanisms underlying the DIP remain unexplored, fundamentally hindering the development of effective mitigation strategies.

In this work, we address this gap by analyzing the geometric relationship between the array manifold and the jamming subspace, revealing two distinct ill-conditioning mechanisms,  namely the observability loss and the manifold ambiguity. 
We find that the estimation accuracy is largely governed by the relative SOI-jamming geometry in the array frame, indicating that array attitude can be leveraged to overcome geometric ill-conditions.
Motivated by this insight, we develop a   \emph{rotational diversity} paradigm, in which the array is rotated to provide geometric diversity for mitigating the DIP. 
Based on this principle, we develop a Rotation-Enhanced SWN-MUSIC framework, which performs independent DOA estimations across multiple predetermined rotations and selectively integrates the reliable estimates. 
Numerical simulations demonstrate that the proposed framework substantially improves estimation accuracy in the problematic directions for fixed-array SWN-MUSIC, without degrading estimation accuracy in other directions.

The main contributions of this work are summarized as follows:
\begin{itemize}
    \item  
    We reveal the physical mechanisms of the DIP by revealing two geometric ill-conditions between the array manifold and the jamming subspace.

    \item  We propose rotational diversity, which elevates array attitude from a passive constraint to a controllable sensing resource for mitigating the DIP.
    
    \item  We develop a  Rotation-Enhanced SWN-MUSIC framework for robust DOA estimation under jamming.
\end{itemize}

The remainder of this paper is structured as follows. Section \ref{sec:problem} formulates the problem and provides a brief review of the SWN-MUSIC algorithm. Section \ref{dip} reveals the underlying mechanisms of the DIP. Section \ref{sec:method} develops the proposed Rotation-Enhanced SWN-MUSIC framework. Section \ref{sec:simulation} presents and analyzes the simulation results. Finally, Section \ref{sec:conclusion} summarizes the work.

Notations: Throughout this paper, vectors and matrices are denoted by boldface lower-case and upper-case characters, respectively, while scalars are represented in lightface. The superscripts $(\cdot)^\mathrm{T}$ and $(\cdot)^\mathrm{H}$ denote the transpose and conjugate transpose, respectively. The operators $(\cdot)^{-1}$, $(\cdot)^{-1/2}$, and $\mathrm{Tr}(\cdot)$ correspond to the matrix inverse, inverse square root, and trace, respectively. The Euclidean norm is denoted by $\|\cdot\|$, while $|\cdot|$ denotes either the absolute value of a scalar or the cardinality of a set, as determined by context.

\section{PRELIMINARIES} \label{sec:problem}

\subsection{Problem Formulation} \label{subsec:Array Signal Model}

\begin{figure}[htbp]
\centering
\includegraphics[width=0.7\linewidth]{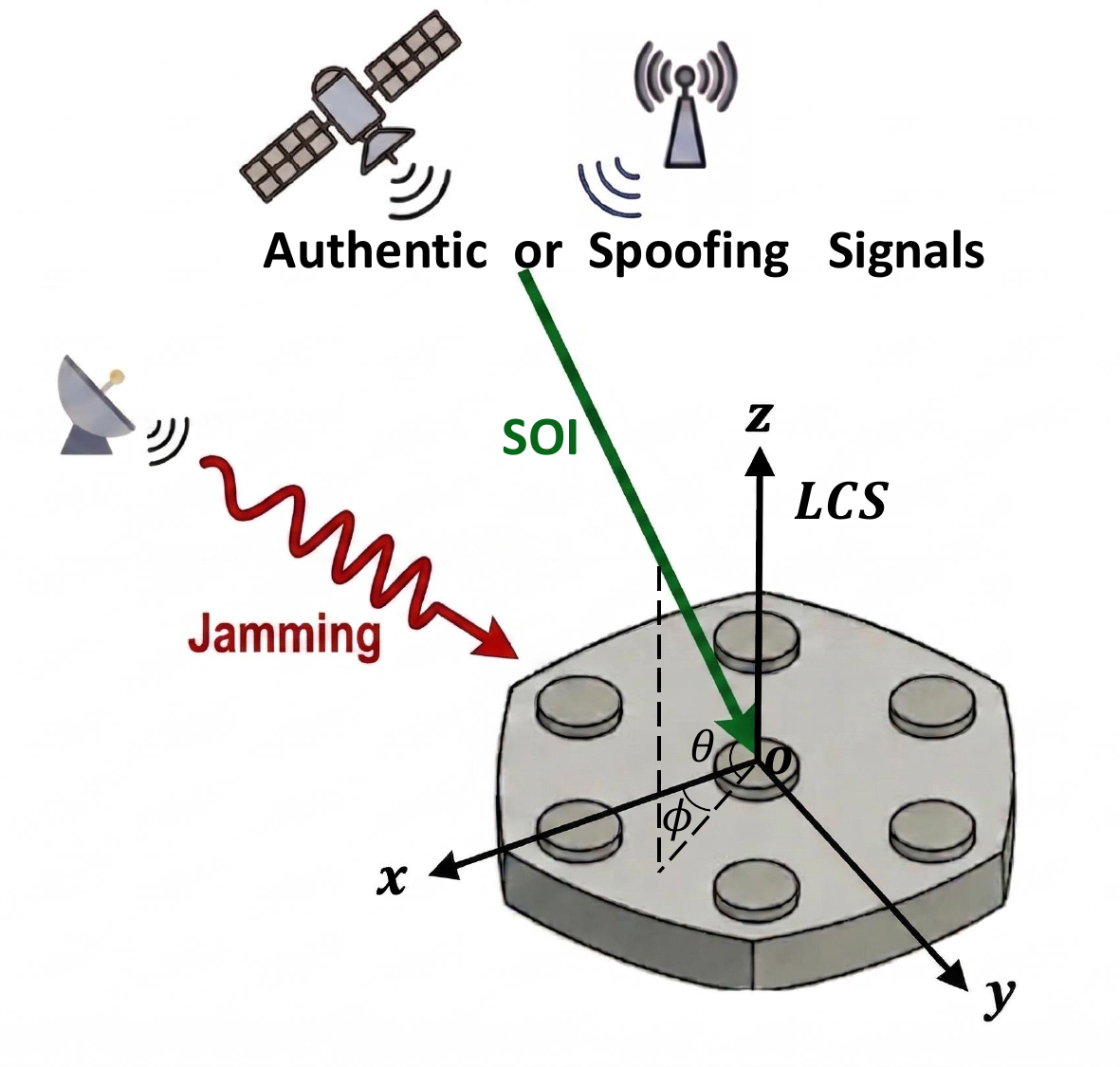}
\caption{Research scenario and coordinate system definitions. }
\label{array}
\end{figure}

Fig.~\ref{array} illustrates the research scenario: an antenna array estimating the DOAs of GNSS-like signals under jamming.
The unknown parameters to be estimated are the SOI azimuth and elevation angles defined in the local coordinate system (LCS), collected as $\boldsymbol{\gamma}=[\phi,\theta]^{\mathrm T}$.

Under the assumptions of the far-field condition and the narrowband signal, the phase differences across the elements are captured by the steering vector:
\begin{equation}
\label{a_simple}
\boldsymbol{a}(\boldsymbol{\gamma}) = \left[e^{-j\frac{2\pi}{\lambda}\boldsymbol{p}_1^\mathrm{T}\boldsymbol{d}(\boldsymbol{\gamma})}, e^{-j\frac{2\pi}{\lambda}\boldsymbol{p}_2^\mathrm{T}\boldsymbol{d}(\boldsymbol{\gamma})}, \ldots, e^{-j\frac{2\pi}{\lambda}\boldsymbol{p}_{M_a}^\mathrm{T}\boldsymbol{d}(\boldsymbol{\gamma})}\right]^\mathrm{T},
\end{equation}
where $\lambda$ is the carrier wavelength, $\boldsymbol{p}_m$ ($m=1,\dots,M_a$) represents the position vector of each element in the LCS.
The mapping function $\boldsymbol{d}(\cdot)$ transforms the DOA into a unit direction vector:
\begin{equation}\label{d_definition}
\boldsymbol{d}([u, v]^\mathrm{T}) =
\begin{bmatrix}
\cos(u)\cos(v) \\
\sin(u)\cos(v) \\
\sin(v)
\end{bmatrix}.
\end{equation}

When  $K$ GNSS-like signals and $K_J$ jamming signals impinge on the array, the sampled baseband signal vector is modeled as
 \begin{multline}
\label{model1}
\boldsymbol{x}(nT_s) = 
\sum_{k=0}^{K-1} \boldsymbol{a}(\boldsymbol{\gamma}_k) s_k(nT_s) 
+ \sum_{k=0}^{K_J-1} \boldsymbol{a}(\boldsymbol{\gamma}_k^J) s_k^J(nT_s) \\
+ \boldsymbol{n}(nT_s),
\end{multline}
where $T_s$ is the sampling interval, $\boldsymbol{\gamma}_k$ and $\boldsymbol{\gamma}_k^J$ are the DOAs of the $k$-th GNSS-like signal and jamming signal, respectively, the terms $s_k(nT_s)$ and $s_k^J(nT_s)$ denote the complex envelopes of the $k$-th GNSS-like signal and jamming signal as received at the LCS origin, and $\boldsymbol{n}$ represents the independent white Gaussian noise.

The $k$-th GNSS-like signal is detailed as
\begin{equation}
\label{GNSS_sig}
s_k(nT_s)=A_kD_k(nT_s-\tau_k)C_k(nT_s-\tau_k)e^{j2\pi(f_\text{IF}+f_k)t+j	
\varphi_k},
\end{equation}
where $A_k, D_k, C_k, \tau_k, f_{\text{IF}}, f_k$, and $\varphi_k$ denote the amplitude, navigation data bit, pseudo-random noise (PRN) code, code delay, intermediate frequency, Doppler frequency, and carrier phase, respectively. Note that the authentic and spoofing GNSS signals are not explicitly distinguished in our signal model, as they share the same signal structure and are processed identically in DOA estimation.

Since this work involves estimating the DOA of the same SOI at different array attitudes, we define the LCS at the array's initial attitude as the fixed Global Coordinate System (GCS). 
DOA estimates obtained in the LCS at any other attitude must be mapped back to this GCS. 
For the sake of narrative continuity, the definition of array attitudes and the coordinate transformations between coordinate systems are provided in the \hyperref[Appendix]{Appendix}.

The objective of this work is to accurately estimate the DOAs 
$\boldsymbol{\gamma}_k$ of GNSS-like signals in a jamming environment,  using the received signal $\boldsymbol{x}(nT_s)$ as the observation, which is modeled by \eqref{model1}.

\subsection{DOA Estimation by SWN-MUSIC}

As a representative state-of-the-art approach, SWN-MUSIC is briefly reviewed to provide context for the subsequent DIP analysis. 
The CRLB of this estimation problem is also summarized, as it will be used in the proposed framework.

As a two-stage algorithm, SWN-MUSIC first estimates the statistical characteristics of the jamming and suppresses it through spatial whitening. 
Since the pre-despreading signal $\boldsymbol{x}$ is dominated by jamming, the covariance matrix of the jamming can be approximated by that of $\boldsymbol{x}$, which is estimated as
\begin{equation}
\label{Rxx}
\hat{\mathbf{R}}_{x} = \frac{1}{N} \sum_{n=0}^{N-1} \boldsymbol{x}(nT_s) \boldsymbol{x}^{\mathrm{H}}(nT_s).
\end{equation}
The spatial whitening matrix  is given by
\begin{equation}
\label{W}
\mathbf{W} = \hat{\mathbf{R}}_{x}^{\,-1/2}.
\end{equation}

Assuming the $k$-th GNSS-like signal is the SOI whose DOA is to be estimated, the resulting array signal after despreading with the corresponding PRN code is expressed as
\begin{equation}
\label{xdes_simple}
\boldsymbol{x}^{\mathrm{des}}_k(nT_c) = \zeta_k\boldsymbol{a}(\boldsymbol{\gamma}_k) + 
\boldsymbol{j}^{\mathrm{des}}(nT_c) + \boldsymbol{n}^{\mathrm{des}}(nT_c),
\end{equation}
where $T_c$ denotes the PRN code period,  $\zeta_k$ is the complex amplitude, 
and $\boldsymbol{j}^{\mathrm{des}}(nT_c)$ and $\boldsymbol{n}^{\mathrm{des}}(nT_c)$ represent the jamming and thermal noise after despreading, respectively.
Applying spatial whitening to the post-despreading signal yields
\begin{equation}
\label{z}
\boldsymbol{z}_k(nT_c) = 
\mathbf{W}\boldsymbol{x}^{\mathrm{des}}_k(nT_c),
\end{equation}
where the residual jamming is approximately spatially white.

The second stage of SWN-MUSIC computes a modified MUSIC pseudospectrum that accounts for the preceding spatial whitening. 
The post-despreading covariance matrix, estimated from $N_c$ snapshots as
\begin{equation}
\label{eq12}
\hat{\mathbf{R}}_{\boldsymbol{z}} = \frac{1}{N_c} \sum_{n=0}^{N_c-1} \boldsymbol{z}(n T_c) \, \boldsymbol{z}^{\mathrm{H}}(n T_c),
\end{equation}
is subsequently decomposed via singular value decomposition (SVD) to extract the noise subspace $\hat{\mathbf{U}}_N$. 
The SWN-MUSIC pseudospectrum is constructed as
\begin{equation}
\label{pseudospectrum}
P_{\text{SWN-MUSIC}}(\boldsymbol{\gamma}) = \frac{\| \mathbf{W}\mathbf{a}(\boldsymbol{\gamma}) \|^2}{\| \hat{\mathbf{U}}_N^H \mathbf{W}\mathbf{a}(\boldsymbol{\gamma}) \|^2}.
\end{equation}
Finally, the DOA estimate is determined by locating the maximum of the pseudospectrum:
\begin{equation}
\label{est_gamma}
\hat{\boldsymbol{\gamma}}_k
=
\arg\max_{\boldsymbol{\gamma}}
P_{\text{SWN-MUSIC}}(\boldsymbol{\gamma}).
\end{equation}

The CRLB for this problem is given by \cite{wang2024two,11231075}:
\begin{equation}
\label{eq:CRLB_DOA}
\begin{aligned}
\mathbf{CRLB}(\boldsymbol{\gamma}_k) &= \frac{1}{2 N_c |\zeta_k|^2} \times \\
&\quad \left\{ \mathrm{Re} \left[ \left( \mathbf{P}_{\mathbf{e}_k}^\perp \frac{\partial \mathbf{e}(\boldsymbol{\gamma}_k)}{\partial \boldsymbol{\gamma}_k} \right)^{\mathrm{H}} \right. \right. \\
&\quad \left. \left. \times \left( \mathbf{P}_{\mathbf{e}_k}^\perp \frac{\partial \mathbf{e}(\boldsymbol{\gamma}_k)}{\partial \boldsymbol{\gamma}_k} \right) \right] \right\}^{-1},
\end{aligned}
\end{equation}
where $\mathbf{e}(\boldsymbol{\gamma}_k) = \mathbf{R}_{x}^{-1/2}\mathbf{a}(\boldsymbol{\gamma}_k)$ and  $\mathbf{P}_{\mathbf{e}_k}^\perp$ denotes the orthogonal projection matrix onto its orthogonal complement. 
Since this $2 \times 2$ matrix does not permit direct scalar comparison,  
a scalar metric \cite{11231075}, $Q(\boldsymbol{\gamma}_k)$, is introduced to characterize the estimation accuracy:
\begin{equation}
\label{Q}
Q(\boldsymbol{\gamma}_k) = \sqrt{ \mathrm{Tr} \left( 
    \begin{bmatrix}
        \cos^2(\theta_k) & 0 \\
        0 & 1
    \end{bmatrix}
    \mathbf{CRLB}(\boldsymbol{\gamma}_k)
\right) }.
\end{equation}
Here, the  factor $\cos^2\theta_k$  accounts for the spherical geometry relating azimuth and elevation errors.

\section{Mechanism Analysis of Direction-Dependent Inaccuracy Phenomenon}
\label{dip}

\begin{figure}[htbp]
\centering
\includegraphics[width=2in]{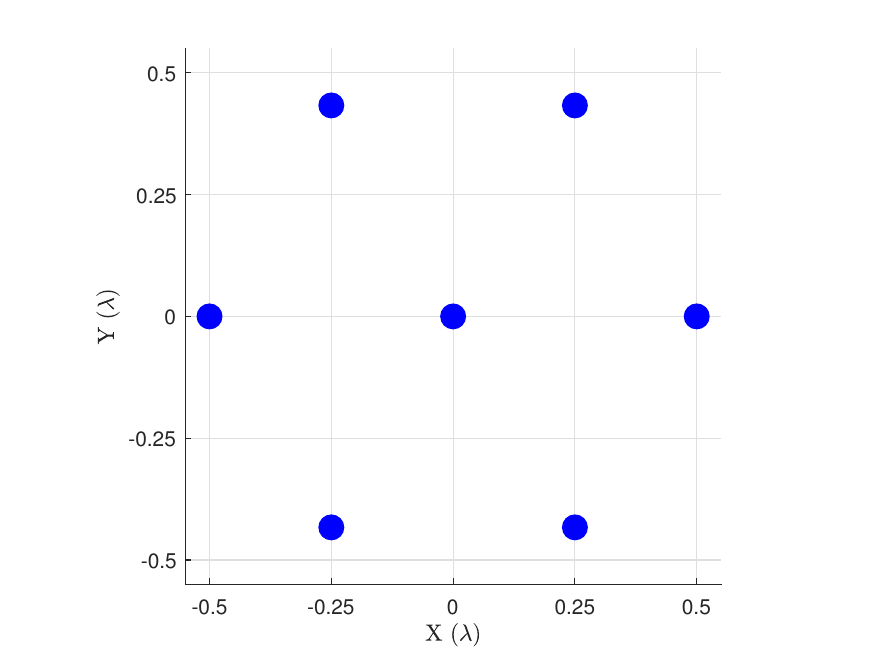}
\caption{Geometry of the 7-element hexagonal antenna array.}
\label{GNSSarray}
\end{figure}

As mentioned earlier, the estimation performance depends on the SOI
direction under a given jamming and array configuration. 
To demonstrate this phenomenon, we first present the CRLB obtained with a seven-element hexagonal array (Fig.~\ref{GNSSarray}), a configuration widely adopted in both research \cite{markel2002interference, wang2024two, 11231075, 11159122}  and commercial products \cite{AntcomCRPA, HarxonAntenna}. 
Other simulation parameters are provided in Table~\ref{tab:simulation_parameters} and Table \ref{tab:interference_config}, where four jammers are configured to simulate a severe electronic attack scenario. 
Throughout this section, the array is fixed at the initial attitude, and all DOAs are defined in the GCS, which coincides with the LCS under this attitude.

\begin{table}[h]
\centering
\caption{Simulation Parameters Configuration}
\label{tab:simulation_parameters}
\begin{tabularx}{\columnwidth}{lX}
\toprule
\textbf{Parameter} & \textbf{Value} \\
\midrule
GNSS Type & GPS L1 C/A \\
Receiver Bandwidth & 5 MHz \\
Sampling Rate & 5 MHz \\
$N_c$ & 30 \\
Jamming-to-Noise ratio (JNR) & 30 dB (per Jammer) \\
Jamming type & Static, wideband Gaussian \\
Jammer DOAs &  Table \ref{tab:interference_config} \\
Signal-to-noise ratio (SNR) & -25 dB\\
\bottomrule
\end{tabularx}
\end{table}

\begin{table}[htbp]
\centering
\caption{Jammer DOAs in the GCS}
\label{tab:interference_config}
\begin{tabular}{ccc}
\toprule
\textbf{Jammer \#} & \textbf{Azimuth ($^\circ$)} & \textbf{Elevation ($^\circ$)} \\
\midrule
1 & 250 & 15 \\
2 & 170  & 30 \\
3 & 100 & 45  \\
4 & 80 & 35 \\
\bottomrule
\end{tabular}
\end{table}

\begin{figure}[H]
\centering
\subfloat[]{\includegraphics[width=0.7\linewidth]{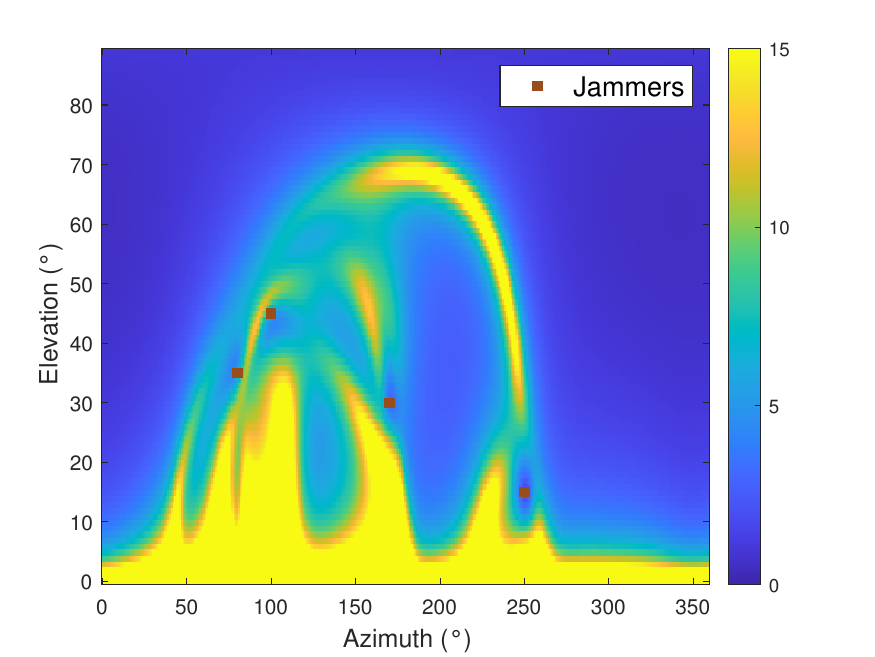}%
\label{figRMSEandCRLB_a}}
\hfil
\subfloat[]{\includegraphics[width=0.7\linewidth]{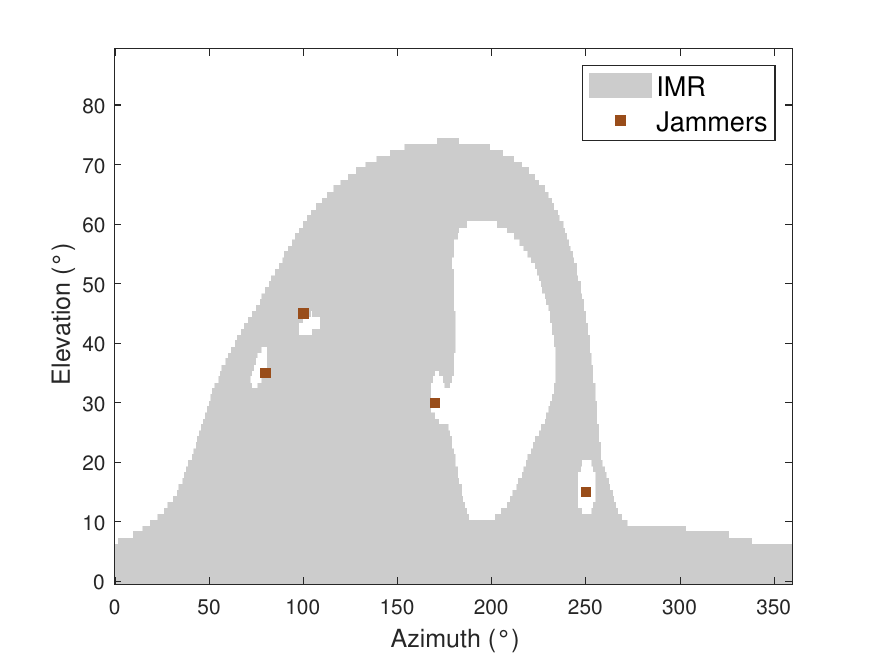}%
\label{figRMSEandCRLB_b}}
\hfil
\caption{DOA estimation accuracy characterized by $Q(\boldsymbol{\gamma})$. (a) Heat map of $Q(\boldsymbol{\gamma})$. (b) Partition of the angular space into the IMR
($Q(\boldsymbol{\gamma})>5^\circ$) and the AMR.}
\label{figRMSEandCRLB}
\end{figure}

As shown in Fig.~\ref{figRMSEandCRLB_a}, the spatial distribution of 
$Q(\boldsymbol{\gamma})$ is highly non-uniform, indicating strong 
direction dependence in DOA estimation accuracy. 
To better illustrate this effect, an accuracy threshold $\eta_Q=5^\circ$ is 
introduced to partition the angular space into the Accurate Measurement 
Region (AMR) and the Inaccurate Measurement Region (IMR). 
Fig.~\ref{figRMSEandCRLB_b} shows that the IMR occupies a substantial 
portion of the angular space, suggesting that SOIs in practice may often 
lie in directions with inherently limited estimation accuracy. 
The localized nulls near the jammers arise from numerical singularities 
and are excluded from analysis, since SOIs in these directions are 
fundamentally unresolvable due to signal masking.

To better understand why directions in the IMR cannot be estimated accurately, several representative examples summarized in Table~\ref{table_summary_regions} are examined next.

\subsection{Mechanism 1: Observability Loss}
 \label{sec:ANALYSIS1}

Our analysis reveals that the DIP arises from two distinct mechanisms.
The first mechanism occurs when the steering vector of the SOI, $\mathbf{a}(\boldsymbol{\gamma})$, falls into the jamming subspace. Formally, this condition is triggered when $\mathbf{a}(\boldsymbol{\gamma})$ can be expressed as a linear combination of the jamming steering vectors \footnote{
\eqref{eq:linear_combination}, and similarly
\eqref{eq:reason2}, should be interpreted in an approximate sense
rather than as strict equalities.
}:
\begin{equation}
\label{eq:linear_combination}
\boldsymbol{a}(\boldsymbol{\gamma}) = \sum_{k=0}^{K_J-1} c_k \boldsymbol{a}(\boldsymbol{\gamma}_k^J).
\end{equation}

\begin{figure}[htbp]
\centering
\includegraphics[width=0.7\linewidth]{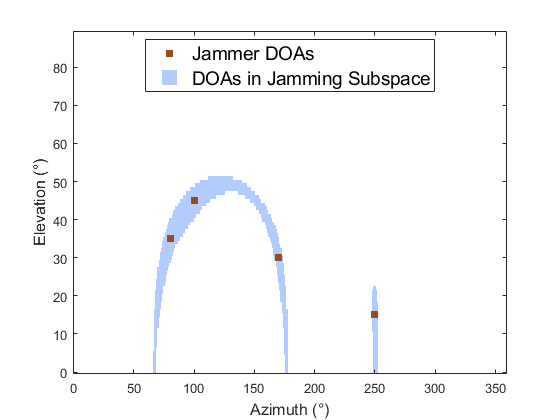}
\caption{Spatial distribution of directions within the jamming subspace (indicated by light-blue regions). }
\label{Jamming_subspace}
\end{figure}

From a physical perspective, this geometric alignment leads to the observability loss.
In jamming environments, nulling the jamming subspace is an inevitable prerequisite for any algorithm to recover an SOI masked by jamming.
Consequently, any SOI whose steering vector lies in this subspace is  suppressed together with the jammers.
Fig.~\ref{Jamming_subspace} illustrates these directions, shown as blue-shaded regions, which constitute a subset of the IMR.

\begin{figure}[htbp]
\centering
\includegraphics[width=0.7\linewidth]{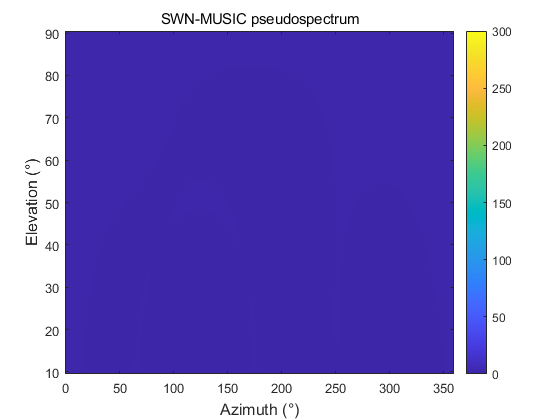}
\caption{SWN-MUSIC pseudospectrum for $\boldsymbol{\gamma} = [125^\circ, 50^\circ]^\mathrm{T}$. The expected peak at $[125^\circ, 50^\circ]^\mathrm{T}$ is lost due to the observability loss. }
\label{pseudospectrum_at_12550}
\end{figure}

\begin{figure}[htbp]
\centering
\includegraphics[width=0.7\linewidth]{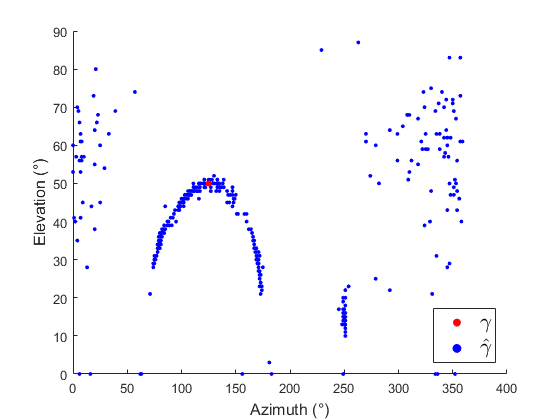}
\caption{Distribution of DOA estimates for $\boldsymbol{\gamma} = [125^\circ, 50^\circ]^\mathrm{T}$.  $\hat{\boldsymbol{\gamma}}$  deviates from $\boldsymbol{\gamma}$ frequently, exhibiting significant errors.}
\label{reason1_sactter}
\end{figure}

In  SWN-MUSIC, when the condition in \eqref{eq:linear_combination} is satisfied, the spatial whitening process compresses the whitened steering vector $\mathbf{W}\mathbf{a}(\boldsymbol{\gamma})$ into a near-zero vector, causing the expected pseudospectrum peak at the true DOA to disappear:
\begin{equation}
\label{eq:collapse}
P_{\text{SWN-MUSIC}}(\boldsymbol{\gamma})
=
\frac{\|\mathbf W\mathbf a(\boldsymbol{\gamma})\|^2}
{\|\hat{\mathbf U}_N^H\mathbf W\mathbf a(\boldsymbol{\gamma})\|^2}
\approx
\frac{0}{0}.
\end{equation}
As an example, $\boldsymbol{\gamma}=[125^\circ,50^\circ]^{\mathrm T}$ is selected to examine the SWN-MUSIC pseudospectrum.
As shown in Fig. \ref{pseudospectrum_at_12550}, no valid peak is exhibited, rendering accurate DOA estimation impossible.
Fig.~\ref{reason1_sactter} further shows the scatter plot of estimates from 500 Monte Carlo trials, exhibiting substantial errors  caused by this observability loss.

\subsection{Mechanism 2: Manifold Ambiguity}
 \label{sec:ANALYSIS2}

Even when $\boldsymbol{\gamma}$ does not lie in the jamming subspace, the DIP can still arise through a second mechanism.
This occurs when a linear combination of the SOI steering vector and jamming steering vectors coincides with a spurious direction:
\begin{equation}
\label{eq:reason2}
c\boldsymbol{a}(\boldsymbol{\gamma}) + \sum_{k=0}^{K_J-1} c_k \boldsymbol{a}(\boldsymbol{\gamma}_k^J) = \boldsymbol{a}(\boldsymbol{\gamma}_{\text{sp}}).
\end{equation}

After jamming suppression removes the jamming-related components, the true DOA 
$\boldsymbol{\gamma}$ and spurious directions $\boldsymbol{\gamma}_{\text{sp}}$
become indistinguishable to the estimator, causing spurious solutions to compete with or replace the true estimate.
From an array-manifold perspective, $\boldsymbol{\gamma}$ and
$\boldsymbol{\gamma}_{\text{sp}}$, originally distinct in the array manifold,
become folded together in the jamming-suppressed manifold, thereby giving rise
to ambiguity.

In SWN-MUSIC, directions satisfying \eqref{eq:reason2} can generate spurious
peaks comparable to the true peak, as we derive
\begin{equation}
\label{eq:indistinguishable}
\begin{aligned}
P_{\text{SWN-MUSIC}}(\boldsymbol{\gamma}_{\text{sp}}) 
&= \frac{\| \mathbf{W}\mathbf{a}(\boldsymbol{\gamma}_{\text{sp}}) \|^2}{\| \hat{\mathbf{U}}_N^H \mathbf{W}\mathbf{a}(\boldsymbol{\gamma}_{\text{sp}}) \|^2} \\
&= \frac{\| c_0\mathbf{W}\mathbf{a}(\boldsymbol{\gamma}) + \mathbf{W}\sum_{j=1}^{K_J} c_j \mathbf{a}_J^j \|^2}{\| c_0\hat{\mathbf{U}}_N^H \mathbf{W}\mathbf{a}(\boldsymbol{\gamma}) + \hat{\mathbf{U}}_N^H\mathbf{W}\sum_{j=1}^{K_J} c_j \mathbf{a}_J^j \|^2} \\
&\approx \frac{\| c_0\mathbf{W}\mathbf{a}(\boldsymbol{\gamma}) \|^2}{\| c_0\hat{\mathbf{U}}_N^H \mathbf{W}\mathbf{a}(\boldsymbol{\gamma}) \|^2} \\
&= \frac{\| \mathbf{W}\mathbf{a}(\boldsymbol{\gamma}) \|^2}{\| \hat{\mathbf{U}}_N^H \mathbf{W}\mathbf{a}(\boldsymbol{\gamma}) \|^2} = P_{\text{SWN-MUSIC}}(\boldsymbol{\gamma}).
\end{aligned}
\end{equation}

\begin{figure}[htbp]
\centering
\includegraphics[width=0.7\linewidth]{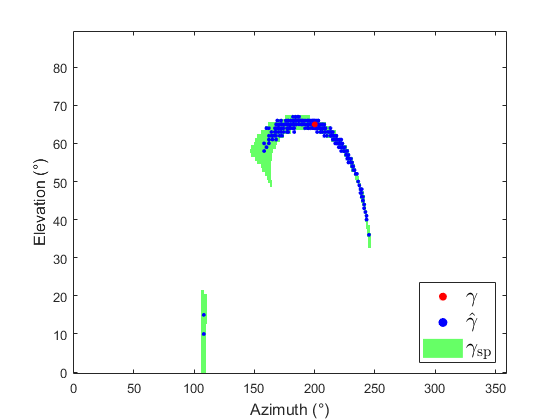}
\caption{ Distribution of DOA estimates for $\boldsymbol{\gamma} = [200^\circ, 65^\circ]^\mathrm{T}$. The light-green regions highlight the spurious directions 
$\boldsymbol{\gamma}_{\text{sp}}$ satisfying \eqref{eq:reason2}, with the blue scatter points showing estimates that align closely, confirming the manifold ambiguity. }
\label{reason2_scatter}
\end{figure}

\begin{figure}[htbp]
\centering
\includegraphics[width=0.7\linewidth]{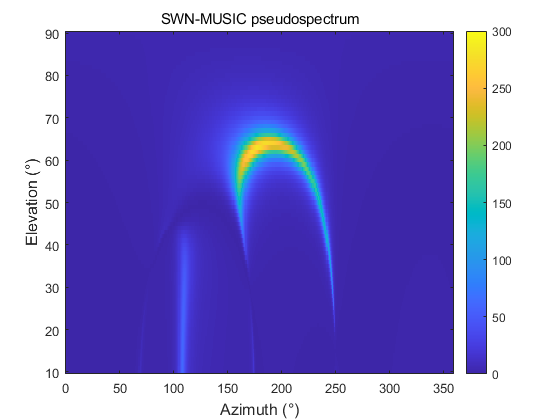}
\caption{SWN-MUSIC pseudospectrum for $\boldsymbol{\gamma} = [200^\circ, 65^\circ]^\mathrm{T}$. Instead of a sharp peak, broadened spectral ridges emerge across the regions of $\boldsymbol{\gamma}_{\text{sp}}$. }
\label{pseudospectrum_at_25065}
\end{figure}

A representative example is
$\boldsymbol{\gamma}=[200^\circ,65^\circ]^{\mathrm T}$,
which exhibits the manifold ambiguity.
As shown in Fig.~\ref{reason2_scatter},  a cluster of spurious directions
$\boldsymbol{\gamma}_{\text{sp}}$ satisfying \eqref{eq:reason2} is highlighted in green,
while the 500 Monte Carlo estimates (blue scatter points) align closely with these spurious directions, confirming that the inaccuracy originates from the manifold ambiguity.
Fig.~\ref{pseudospectrum_at_25065} shows the SWN-MUSIC pseudospectrum from a single experiment, where broadened spectral ridges appear over the $\boldsymbol{\gamma}_{\text{sp}}$ regions rather than a sharp peak at the true DOA, providing additional evidence of this ambiguity.

\subsection{Summary of DIP Mechanism}

The two mechanisms are not mutually exclusive.
Indeed, \eqref{eq:linear_combination} can be viewed as a special case of
\eqref{eq:reason2}, where $\boldsymbol{\gamma}_{\mathrm{sp}}$
degenerates into directions within the jamming subspace.
This also explains Fig.~\ref{reason1_sactter}, where part of the estimates forms bands aligned with the jamming subspace, while others scatter stochastically under noise dominance caused by severe SOI suppression.

\begin{figure}[htbp]
\centering
\includegraphics[width=0.8\linewidth]{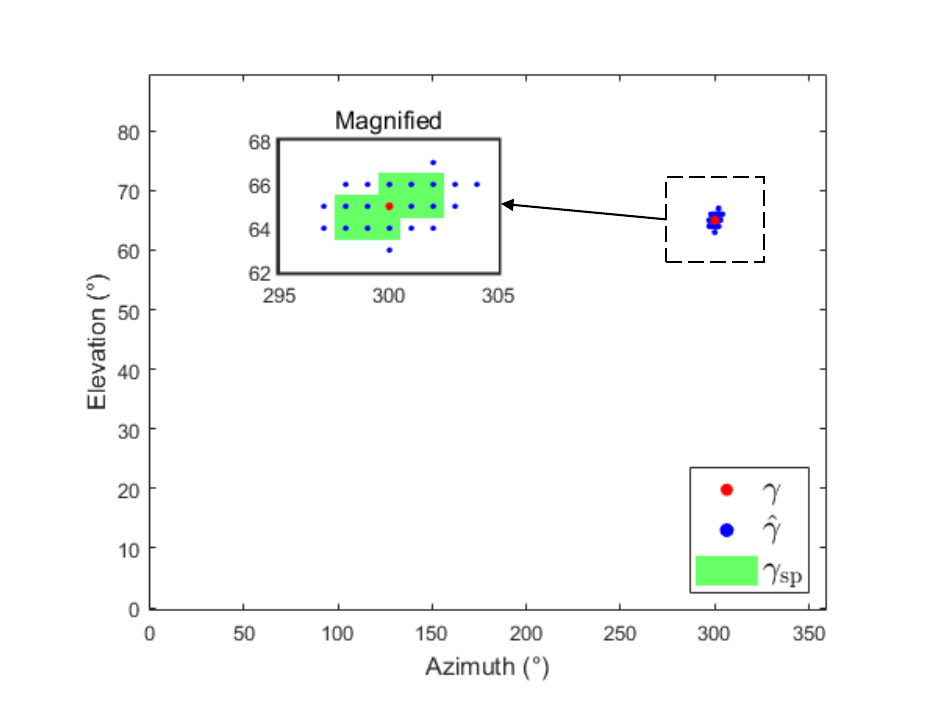}
\caption{ Distribution of DOA estimates for $\boldsymbol{\gamma} = [300^\circ, 65^\circ]^\mathrm{T}$.  The spurious directions $\boldsymbol{\gamma}_{\text{sp}}$ satisfying
\eqref{eq:reason2} are confined to a small neighborhood of the true DOA,
resulting in no significant spatial ambiguity.}
\label{Magnified}
\end{figure}

\begin{figure}[htbp]
\centering
\includegraphics[width=0.7\linewidth]{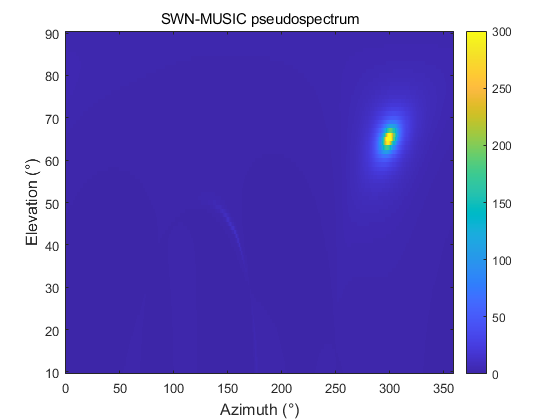}
\caption{SWN-MUSIC pseudospectrum for $\boldsymbol{\gamma} = [300^\circ, 65^\circ]^\mathrm{T}$, showing clearly resolved peaks, as neither of the two mechanisms is triggered.} 
\label{pseudospectrum_at_30065}
\end{figure}

Accurate DOA estimation is attainable only when neither mechanism is triggered,
as exemplified by $\boldsymbol{\gamma}=[300^\circ,65^\circ]^{\mathrm T}$.
This direction lies outside the jamming subspace, while the associated
$\boldsymbol{\gamma}_{\text{sp}}$ appears only in an infinitesimal neighborhood of the true DOA (Fig.~\ref{Magnified}), insufficient to trigger the second mechanism.
Consequently, the estimates cluster tightly around the true DOA without spatial ambiguity, as further evidenced by the sharp pseudospectrum peak in
Fig.~\ref{pseudospectrum_at_30065}.

In summary, both the observability loss and the manifold ambiguity originate from the
geometric relationship between the array manifold and the jamming subspace.
Specifically, the relative geometry between the SOI and the jamming directions in the LCS plays a crucial role in the accuracy of DOA estimation.
This insight motivates us to develop a new method, which improves accuracy by actively changing the geometric relationship, in the next section.

\begin{table}[htbp]
\centering
\renewcommand{\arraystretch}{1.2} 
\setlength{\tabcolsep}{6pt} % 进一步缩小列间距
\caption{Characteristics and Estimation Performance of Representative SOI Directions}
\label{table_summary_regions}
\begin{tabular}{ccccc} 
\toprule
\textbf{DOA} $\boldsymbol{\gamma}$ ($^\circ$) & $Q(\boldsymbol{\gamma})$ & \textbf{Region} & \textbf{Mechanism} & \textbf{RMSE} ($^\circ$) \\
\midrule
$[125, 50]^\mathrm{T}$ & 7.57 & IMR & 1 & 56.05 \\
$[200, 65]^\mathrm{T}$ & 10.33 & IMR & 2 & 15.14 \\
$[300, 65]^\mathrm{T}$ & 0.69 & AMR & --- & 0.73 \\
\bottomrule
\end{tabular}
\end{table}

\section{PROPOSED METHOD} \label{sec:method}

Motivated by the geometric interpretation of the DIP, we develop a Rotation-Enhanced SWN-MUSIC framework. 
Under a fixed array-jammer geometry, the CRLB establishes the fundamental limit of estimation, 
and algorithmic refinement alone cannot overcome the resulting geometry ill-conditions. 
Consequently, an inevitable way to break this limitation is to reconfigure the observation geometry.

To this end, we introduce rotational diversity, which leverages array attitude as a controllable sensing resource to provide multiple observational perspectives. 
The following subsections first establish the principle of rotational diversity and then present the implementation of the proposed framework.

\subsection{Rotational Diversity}
\label{subsec:concept}

For a given direction $\boldsymbol{\gamma}$ defined in the GCS, its direction expressed in the LCS depends on the array attitude. 
According to the coordinate transformation in the \hyperref[Appendix]{Appendix}, the steering vector in the LCS can be written as
\begin{equation}
\label{a_attitude}
\boldsymbol{a}_{\boldsymbol{A}}(\boldsymbol{\gamma}) =
\left[
e^{-j\frac{2\pi}{\lambda}\boldsymbol{p}_1^\mathrm{T}\boldsymbol{R}(\boldsymbol{A})^\mathrm{T}\boldsymbol{d}(\boldsymbol{\gamma})},
\ldots,
e^{-j\frac{2\pi}{\lambda}\boldsymbol{p}_{M_a}^\mathrm{T}\boldsymbol{R}(\boldsymbol{A})^\mathrm{T}\boldsymbol{d}(\boldsymbol{\gamma})}
\right]^\mathrm{T},
\end{equation}
which reduces to \eqref{a_simple} when $\boldsymbol{A} = [0^\circ, 0^\circ, 0^\circ]^\mathrm{T}$.

As a result, the array attitude effectively reparameterizes the array manifold, causing both the SOI steering vector and the jamming subspace to vary with $\boldsymbol{A}$. 
Therefore, the estimation performance of a given direction $\boldsymbol{\gamma}$ can be viewed as  a pseudo-function
\begin{equation}
Q(\boldsymbol{\gamma}) = F(\boldsymbol{\gamma}, \mathcal{J}, \boldsymbol{P}, \boldsymbol{A}),
\end{equation}
where $\mathcal{J}$ denotes the jamming configuration, and $\boldsymbol{P} = \{\boldsymbol{p}_m\}_{m=1}^{M_a}$ specifies the array topology via the element positions, while $\boldsymbol{\gamma}$ and $\boldsymbol{A}$ have been defined previously.
This formulation highlights that the estimation performance is fundamentally governed by the array attitude. 
Together with the other variables, it defines the observation geometry, within which $\boldsymbol{A}$ is the only controllable degree of freedom, making it the primary means for performance modulation.

To reconfigure the observation geometry, a practical and tractable form of attitude adjustment is required. 
Inspired by the antenna rotation method from \cite{sun2022anti}, we restrict the attitude adjustment to in-plane rotation about the array's boresight, i.e., varying only the yaw angle while keeping pitch and roll constant. 
This choice offers two key advantages: it simplifies implementation by requiring only a single-axis rotation mechanism without additional swept volume.
In addition, it streamlines the coordinate transformation, as the DOA in the GCS can be recovered by a simple azimuth compensation while the elevation remains unchanged (see \hyperref[Appendix]{Appendix}).

% To adjust the observation geometry, a practical and tractable form of attitude adjustment is required. 
% In this study, the attitude adjustment is restricted to in-plane rotation about the array's boresight, i.e., varying only the yaw angle while keeping pitch and roll constant, which is sufficient to induce meaningful variations in the observation geometry. 
% This choice offers two key advantages: it simplifies implementation by requiring only a single-axis rotation mechanism without additional swept volume, while similar rotation-assisted GNSS array systems have also been explored in prior anti-jamming studies \cite{sun2022anti}.
% In addition, it streamlines the coordinate transformation, as the DOA in the GCS can be recovered by a simple azimuth compensation while the elevation remains unchanged (see \hyperref[Appendix]{Appendix}).

\begin{figure}[htbp]
\centering
\includegraphics[width=0.75\linewidth]{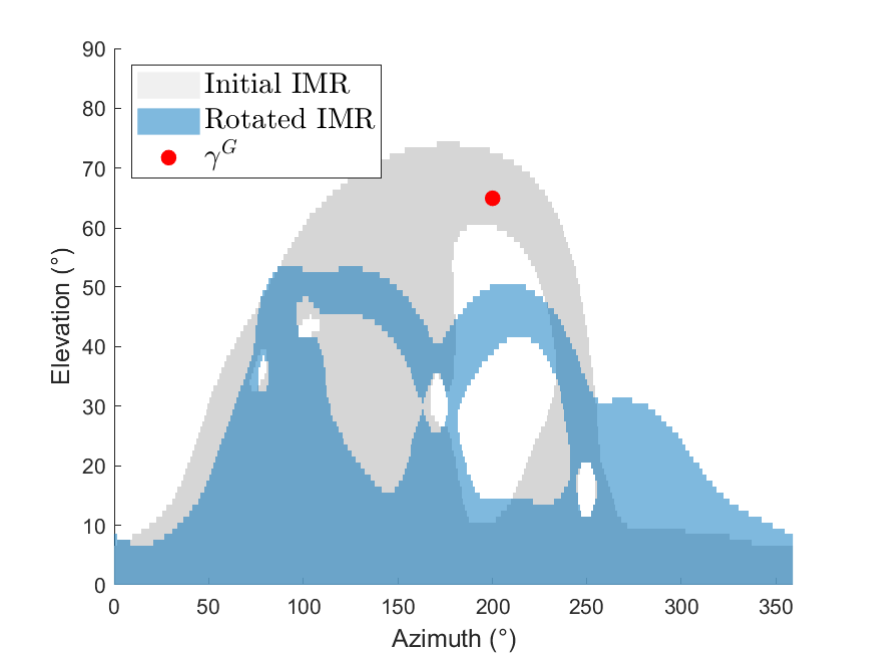}
\caption{IMR and AMR distributions under different array attitudes. The direction $\boldsymbol{\gamma} = [200^\circ, 65^\circ]^\mathrm{T}$ lies within the IMR at $\boldsymbol{A} = [0^\circ, 0^\circ, 0^\circ]^\mathrm{T}$ (grey region) and is relocated into the AMR at $\boldsymbol{A} = [20^\circ, 0^\circ, 0^\circ]^\mathrm{T}$ (blue region), illustrating the geometry reconfiguration induced by array rotation.}
\label{IMR_overlap}
\end{figure}

\begin{figure}[htbp]
\centering
\includegraphics[width=0.8\linewidth]{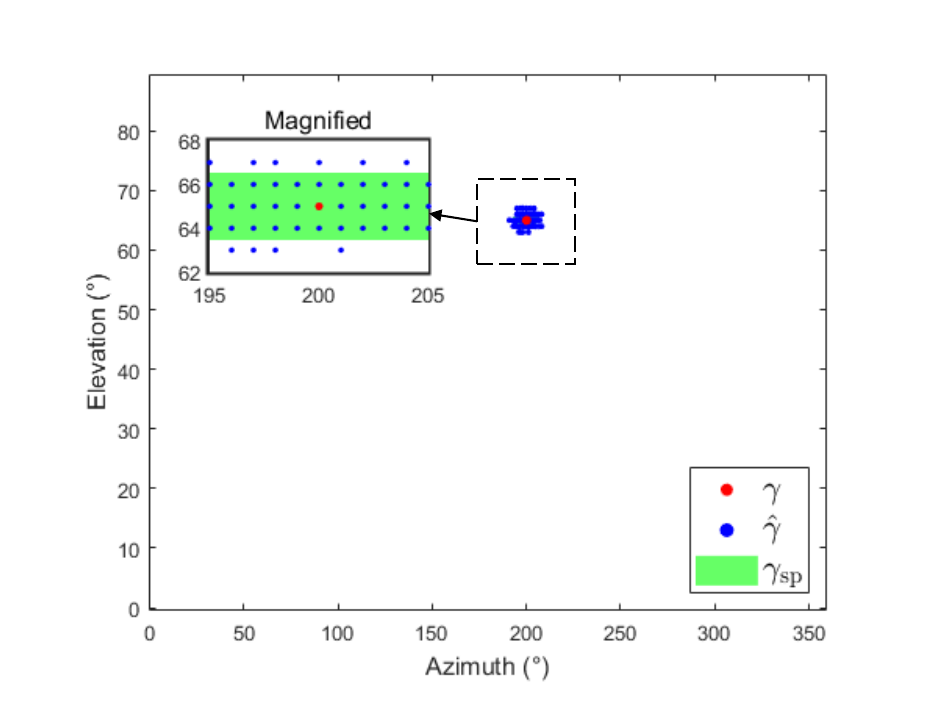}
\caption{Distribution of DOA estimates for $\boldsymbol{\gamma} = [200^\circ, 65^\circ]^\mathrm{T}$ in the GCS after a $20^\circ$ array rotation. Compared with Fig.~\ref{reason2_scatter}, the estimation error is significantly reduced, and Mechanism~2 is no longer triggered.}
\label{scatter_after_rotation}
\end{figure}

To illustrate the effect of array rotation on the estimation performance, we revisit the example $\boldsymbol{\gamma} = [200^\circ, 65^\circ]^\mathrm{T}$, which suffers from significant errors due to Mechanism~2 under the initial attitude. 
As shown in Fig.~\ref{IMR_overlap}, this direction lies within the IMR at $\boldsymbol{A} = [0^\circ, 0^\circ, 0^\circ]^\mathrm{T}$, whereas a change to $\boldsymbol{A} = [20^\circ, 0^\circ, 0^\circ]^\mathrm{T}$ reshapes the CRLB distribution and relocates it into the AMR. 
Consequently, the $Q$ value drops from $10.33$ to $1.33$. 
This improvement is further confirmed in Fig.~\ref{scatter_after_rotation}, where the root mean square error (RMSE), defined by the angular separation between the estimated and true direction vectors, decreases from $15.14^\circ$ to $1.41^\circ$ and Mechanism~2 is no longer triggered.

Since the incident direction of the SOI is unknown, it is not possible to determine a priori which array attitude yields a favorable observation geometry. 
This necessitates the use of multiple predetermined attitudes, for example by rotating the array over a full cycle with uniform angular increments. 
This approach exploits geometric complementarity, whereby an SOI residing in the IMR under one attitude can be possibly in the AMR under another.

\begin{figure}[htbp]
\centering
\includegraphics[width=\columnwidth]{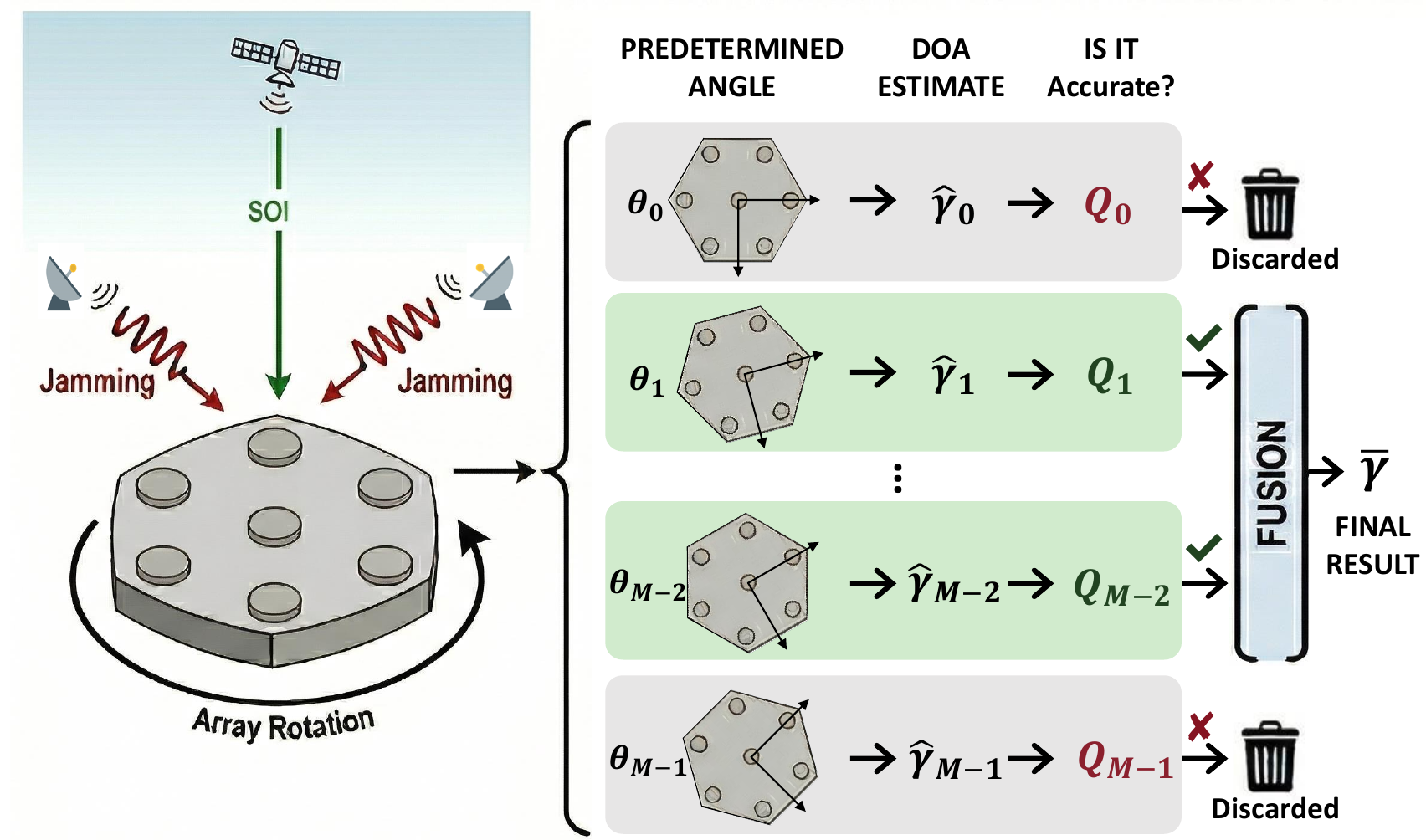}
\caption{Conceptual illustration of rotational diversity. }
\label{rotational_diversity}
\end{figure}

As a result, multiple DOA estimates are obtained, among which accurate and unreliable ones may coexist. 
To extract a reliable final result, a reliability-aware selection and fusion mechanism is adopted, as detailed in the next subsection. 
This leads to a diversity-based perspective in which array rotation is exploited to actively reconfigure the observation geometry and mitigate the ill-condition inherent in a single observation configuration.
We term this approach \emph{rotational diversity}, as illustrated in Fig.~\ref{rotational_diversity}.

\subsection{Rotation-Enhanced SWN-MUSIC Framework}
\label{section_known}

Based on the rotational diversity principle, we develop the Rotation-Enhanced SWN-MUSIC framework, as summarized in \hyperref[alg:rotation_enhanced]{Algorithm}.
The framework operates under two practical assumptions: a step-and-stare acquisition protocol, where data are collected when the array is momentarily stationary to avoid Doppler effects, and a quasi-static SOI direction. Given the short data collection duration, the angular displacement of GNSS satellites is negligible, allowing their DOAs to be treated as constant.

% \captionsetup[algorithm]{labelformat=empty} % 去掉 "Algorithm 1:" 这种前缀
\begin{algorithm}[t]
\renewcommand{\thealgorithm}{}
\caption{Rotation-Enhanced SWN-MUSIC Framework}
\label{alg:rotation_enhanced}
\begin{algorithmic}[1]
\REQUIRE Satellite set $\mathcal{S}$, predefined rotation angles $\{\theta_{m}\}_{m=0}^{M-1}$, threshold $\eta_{Q}$
\ENSURE Final DOA estimates $\bar{\gamma}_{k}$ for each $s_{k} \in \mathcal{S}$
\STATE Initialize $\mathcal{M}_{k} \leftarrow \emptyset$ for each $s_{k} \in \mathcal{S}$
\FOR{each rotation angle $\theta_{m}$ ($m=0, \dots, M-1$)}
    \STATE Rotate the antenna array to $\theta_{m}$ \COMMENT{$\theta_{0}$ is the initial attitude}
    \FOR{each satellite $s_{k} \in \mathcal{S}$}
        \STATE Estimate DOA $\hat{\gamma}_{m,k}$ via SWN-MUSIC
        \STATE Evaluate reliability $Q_{m,k} \leftarrow Q(\hat{\gamma}_{m,k})$ using \eqref{Q}
        \IF{$Q_{m,k} < \eta_{Q}$}
            \STATE $\hat{\gamma}_{m,k} \leftarrow \hat{\gamma}_{m,k} + \theta_{m}$ \COMMENT{Coordinate transformation}
            \STATE $\mathcal{M}_{k} \leftarrow \mathcal{M}_{k} \cup \{\hat{\gamma}_{m,k}\}$
        \ENDIF
    \ENDFOR
\ENDFOR
\FOR{each satellite $s_{k} \in \mathcal{S}$}
    \IF{$\mathcal{M}_{k} \neq \emptyset$}
        \STATE $\bar{\gamma}_{k} \leftarrow \text{fuse\_estimates}(\mathcal{M}_{k})$ 
        \COMMENT{\eqref{strategy1}, \eqref{strategy2}, or \eqref{strategy3} }
        
    \ELSE
        \STATE $\bar{\gamma}_{k} \leftarrow \text{None}$ \COMMENT{Inaccurate estimates in all attitudes}
    \ENDIF
\ENDFOR
\RETURN $\{\bar{\gamma}_{k}\}$
\end{algorithmic}
\end{algorithm}

Line~6 employs $Q$ as a reliability-aware metric, where only estimates with sufficiently low $Q$ values are retained for fusion, while unreliable ones are discarded (Lines~7--10). Since the computation of $Q$ requires the unknown true DOA $\boldsymbol{\gamma}$, it is evaluated using the estimate $\hat{\boldsymbol{\gamma}}$ instead. We adopt this approximation from prior work~\cite{11231075}, which is found to be effective in distinguishing reliable and unreliable estimates in practice.

The signal amplitude $\zeta_k$ required in $Q$ is estimated following the maximum likelihood approach in~\cite{markel2002interference}. As amplitude estimation is not the focus of this paper, the implementation details are omitted for brevity.

Line~15 describes the fusion of all qualified estimates to yield a final result, leveraging the fact that accurate DOA estimates may be obtained from multiple array attitudes. Formally, the set of qualified estimates is defined as $\mathcal{M}_k = \{ \hat{\gamma}_{m,k} \}_{m \in \mathcal{I}_k}$, where $\mathcal{I}_k = \{ m \mid Q_{m,k} \leq \eta_Q \}$. 
To derive the final estimate $\bar{\gamma}_k$ from $\mathcal{M}_k$, three representative fusion strategies are considered, ranging from selection-based to averaging-based approaches:

\begin{enumerate}
    \item \textbf{Min-Q Selection}: This strategy selects the estimate associated with the minimum $Q$ value:
    \begin{equation}
    \label{strategy1}
        \bar{\gamma}_k = \hat{\gamma}_{m,k}, \quad \text{where } m = \arg \min_{i \in \mathcal{I}_k} Q_{i,k}.
    \end{equation}
    
    \item \textbf{Arithmetic Averaging}: This strategy computes the arithmetic mean of all qualified estimates:
    \begin{equation}
    \label{strategy2}
        \bar{\gamma}_k = \frac{1}{|\mathcal{I}_k|} \sum_{m \in \mathcal{I}_k} \hat{\gamma}_{m,k}.
    \end{equation}
    
    \item \textbf{Weighted Averaging}: This strategy performs a weighted average, where each estimate is assigned a weight inversely proportional to its $Q$ value:
    \begin{equation}
    \label{strategy3}
        \bar{\gamma}_k = \sum_{m \in \mathcal{I}_k} w_{m,k} \hat{\gamma}_{m,k}, \quad 
        w_{m,k} = \frac{Q_{m,k}^{-1}}{\sum_{i \in \mathcal{I}_k} Q_{i,k}^{-1}}.
    \end{equation}
\end{enumerate}

The fusion is performed in the direction vector domain before being converted back to angular estimates, which ensures a physically consistent combination compared to direct angle-domain averaging.

\section{Simulation Results} \label{sec:simulation}

In this section, MATLAB-based simulations are conducted to evaluate the proposed Rotation-Enhanced SWN-MUSIC framework. 
The simulations aim to evaluate the estimation accuracy and to illustrate how array rotation reconfigures the observation geometry and expands the reliable angular coverage.
Due to the potential risks and legal restrictions associated with real-world GNSS spoofing and jamming, high-fidelity simulation provides a practical basis for performance verification.

\subsection{Simulation Setup}

\begin{table}[htbp]
\centering
\caption{Satellite DOAs in the GCS}
\label{tab:Satellite_Configuration}
\begin{tabular}{ccc}
\toprule
\textbf{PRN \#} & \textbf{Azimuth ($^\circ$)} & \textbf{Elevation ($^\circ$)} \\
\midrule
1 & 300 & 80 \\
2 & 210  & 30 \\
3 & 160 & 55  \\
4 & 50 & 50 \\
5 & 200 & 65 \\
\bottomrule
\end{tabular}
\end{table}

\begin{figure}[h]
\centering
\includegraphics[width=2in]{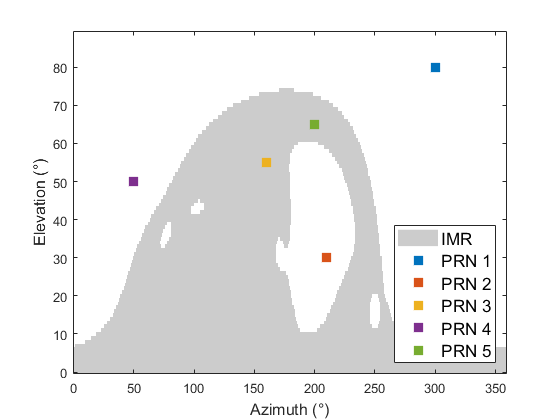}
\caption{Spatial distribution of the five test satellites relative to the IMR, in the GCS.}
\label{sat_distribution}
\end{figure}

Five GNSS signals with PRN codes 1--5 are considered, with their DOAs specified in Table~\ref{tab:Satellite_Configuration}. This setup reflects a typical multi-satellite observation scenario. 
No distinction is made between authentic and spoofing signals, as the focus is on DOA estimation.

The jammer configuration and array geometry are kept consistent with those described in Section~\ref{sec:problem}. 
Fig.~\ref{sat_distribution} illustrates the resulting satellite distribution, with the IMR shaded in grey.

The initial array attitude is set to $\boldsymbol{A} = [0^\circ, 0^\circ, 0^\circ]^\mathrm{T}$, at which SWN-MUSIC is executed as a reference.
The proposed framework starts from the same attitude and then samples additional rotated attitudes.

\subsection{Effectiveness Validation}

\begin{figure*}[t]
\centering
\begin{subfigure}[t]{0.32\textwidth}
    \centering
    \includegraphics[width=\linewidth]{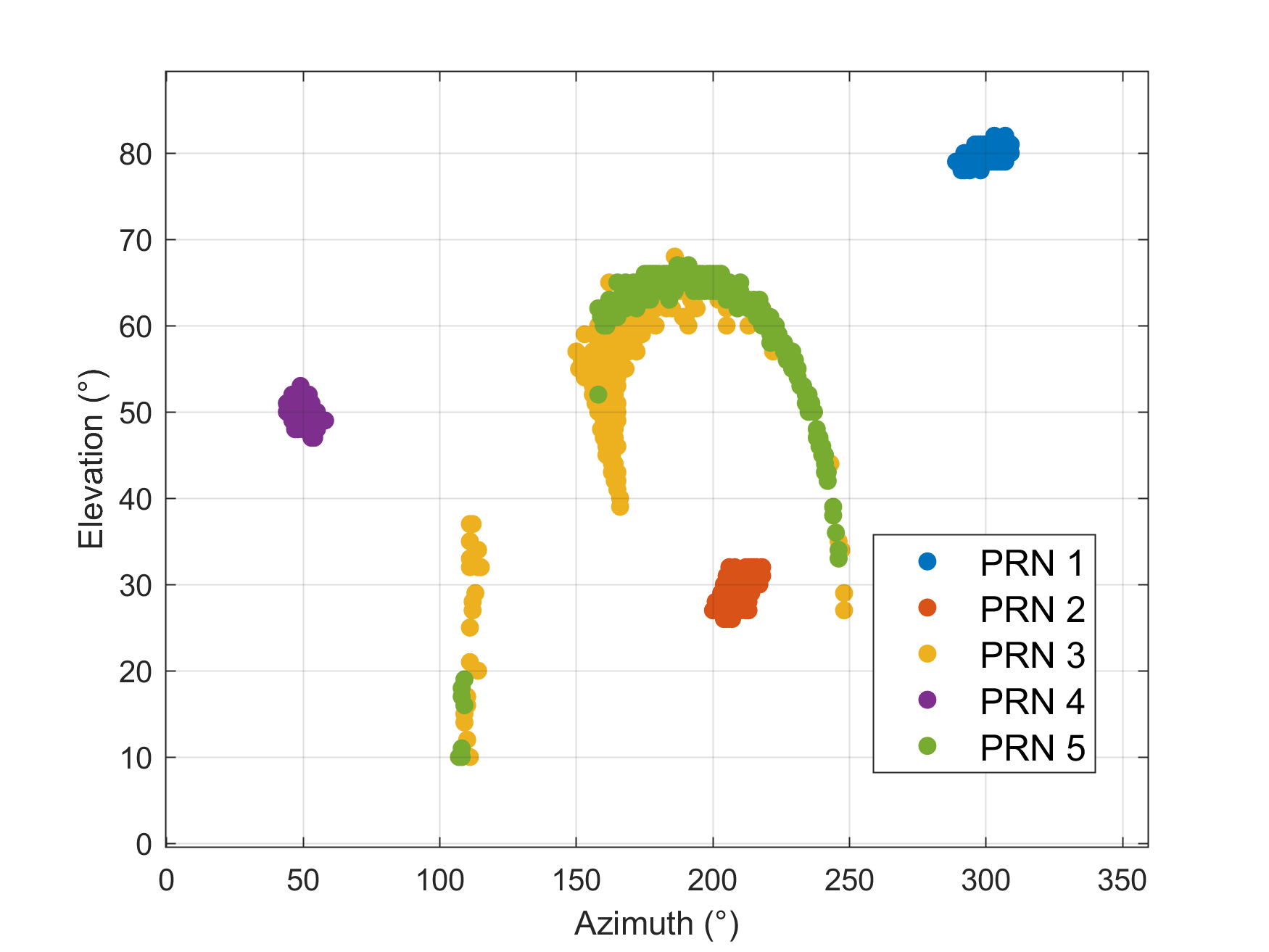}
    \caption{}
    \label{no_rot_a}
\end{subfigure}
\hfill
\begin{subfigure}[t]{0.32\textwidth}
    \centering
    \includegraphics[width=\linewidth]{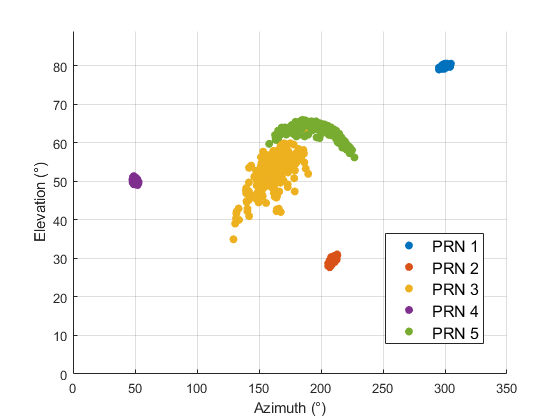}
    \caption{}
    \label{no_rot_b}
\end{subfigure}
\hfill
\begin{subfigure}[t]{0.32\textwidth}
    \centering
    \includegraphics[width=\linewidth]{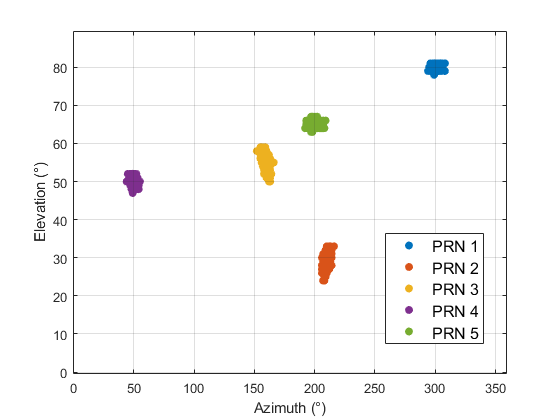}
    \caption{}
    \label{no_rot_c}
\end{subfigure}

\caption{DOA estimation scatter plots from 500 Monte Carlo trials: 
(a) SWN-MUSIC at the initial attitude; 
(b) repeated SWN-MUSIC without rotation (six runs averaged); 
(c) proposed Rotation-Enhanced framework. 
Satellites 3 and 5 exhibit significant estimation errors in (a) and (b), while accurate estimates are achieved in (c).}
\label{no_rot}
\end{figure*}

\begin{table}[htbp]
\centering
\scriptsize % 或 \footnotesize
\caption{DOA Estimation RMSE ($^\circ$)}
\label{tab:RMSE_comparison}
\begin{tabular}{lccccc}
\toprule
Method & PRN 1 & PRN 2 & PRN 3 & PRN 4 & PRN 5 \\
\midrule
SWN-MUSIC & 0.91 & 3.15 & 16.25 & 1.68 & 16.01 \\
Repeated SWN-MUSIC & 0.41 & 1.50 & 7.19 & 0.68 & 5.76 \\
Proposed & 0.67 & 2.49 & \textbf{1.96} & 1.54 & \textbf{1.41} \\
\bottomrule
\end{tabular}
\end{table}

Fig.~\ref{no_rot} presents the DOA estimation scatter plots from 500 Monte Carlo trials. As shown in Fig.~\ref{no_rot_a}, SWN-MUSIC at the initial attitude yields accurate estimates for three satellites, while Satellites~3 and~5 exhibit significant errors, as their true DOAs lie within the IMR.

Fig.~\ref{no_rot_c} presents the results of the proposed framework. 
The Min-Q selection strategy is applied across six sampled attitudes, $\theta_m \in \{0^{\circ}, 10^{\circ}, \dots, 50^{\circ}\}$, corresponding to a $10^{\circ}$ rotation increment from the initial state. 
Sampling stops at $50^{\circ}$ due to the six-fold rotational symmetry of the hexagonal array, under which the array manifold repeats at $60^{\circ}$. 
As summarized in Table~\ref{tab:RMSE_comparison}, the estimation accuracy for Satellites~3 and~5 is significantly improved, while the high accuracy of the other three satellites is consistently maintained.

To verify that the performance gain does not stem solely from the increased number of estimates, Fig.~\ref{no_rot_b} presents a reference case in which SWN-MUSIC is executed six times at the initial attitude, matching the number of attitudes used in the proposed framework, and the resulting estimates are averaged. 
Compared with Fig.~\ref{no_rot_a}, the estimates for Satellites~3 and~5 show a clear reduction in dispersion, yet remain widely scattered and thus unreliable. 
This observation suggests that repeated estimation can provide limited refinement but is insufficient to resolve the fundamentally unfavorable observation geometries  associated with these directions.

\begin{figure*}[htbp] % [htbp] 是浮动体位置参数
    \centering
    \includegraphics[width=\textwidth]{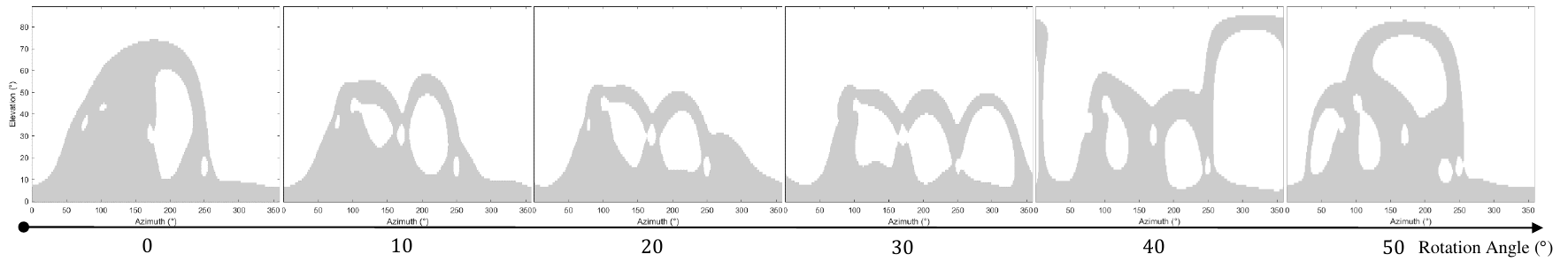} % width=\textwidth 使其跨栏
    \caption{Redistribution of AMR and IMR under array rotation, in the GCS.}
    \label{figQ}
\end{figure*}

To reveal the mechanism behind the accuracy improvement, we show the redistribution of IMR and AMR under different rotation angles in Fig.~\ref{figQ}. 
Based on Fig.~\ref{figQ}, the status of each satellite in the AMR/IMR is summarized in Table~\ref{tab:Visibility_Status}, where `A' and `I' indicate whether a satellite lies in the AMR or IMR, respectively. 
It can be observed that, due to the changing IMR/AMR distribution, Satellites 3 and 5, initially within the IMR under the initial attitude, enter the AMR at least once across the sampled attitudes, ensuring their DOAs are observed under at least one favorable geometry for accurate DOA estimation.

% To reveal the mechanism behind the accuracy improvement, we examine how the IMR and AMR are redistributed under different array attitudes, as illustrated in Fig.~\ref{figQ}. 
% In particular, Satellites~3 and~5, which are initially trapped in the IMR under the initial attitude, can be remapped into the AMR through array rotation. Table~\ref{tab:Visibility_Status} summarizes this behavior, where `A' and `I' indicate whether a satellite lies in the AMR or IMR, respectively. 
% It can be observed that each satellite enters the AMR at least once across the sampled attitudes, thereby ensuring that their DOAs are observed under at least one favorable geometry for accurate DOA estimation.

\begin{table}[htbp]
\centering
\caption{Status (AMR/IMR) of Satellites across Different Rotation Angles}
\label{tab:Visibility_Status}
\begin{tabular}{ccccccc}
\toprule
\textbf{PRN} \# & \textbf{0°} & \textbf{10°} & \textbf{20°} & \textbf{30°} & \textbf{40°} & \textbf{50°} \\ \midrule
1            & A           & A            & A            & A            & I            & A            \\
2            & A           & A            & A            & A            & A            & I            \\
3            & I           & A            & A            & A            & A            & A            \\
4            & A           & A            & A            & A            & A            & I            \\
5            & I           & A            & A            & A            & A            & A            \\ \bottomrule
\end{tabular}
\end{table}

\begin{figure}[htbp]
\centering
% 增加宽度，减小间距
\begin{subfigure}[t]{0.65\columnwidth}  % 增加宽度
    \centering
    \includegraphics[width=\linewidth]{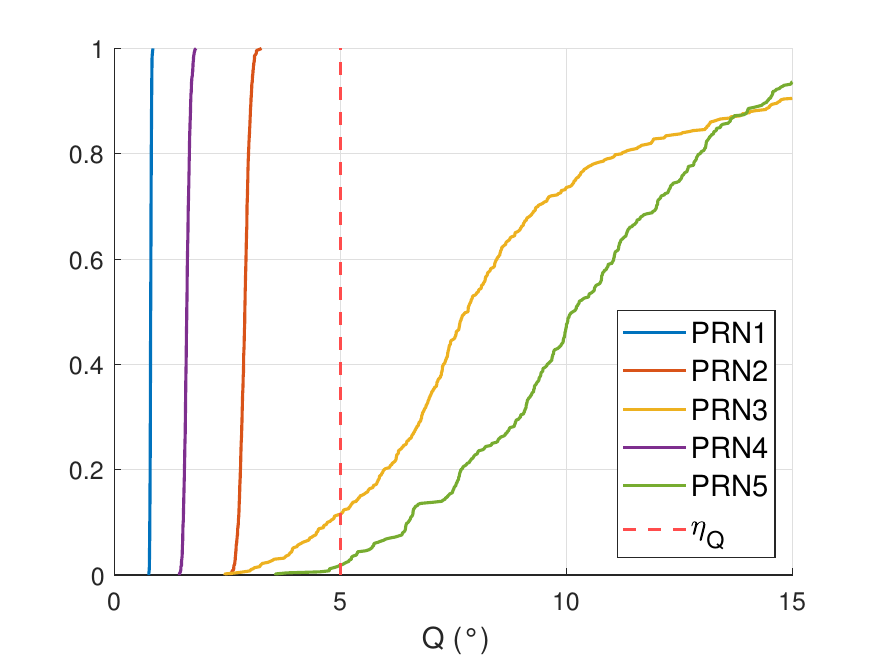}
    \caption{}
    \label{Q_cdf}
\end{subfigure}
\hspace{0\columnwidth}  % 减小间距
\begin{subfigure}[t]{0.65\columnwidth}  % 增加宽度
    \centering
    \includegraphics[width=\linewidth]{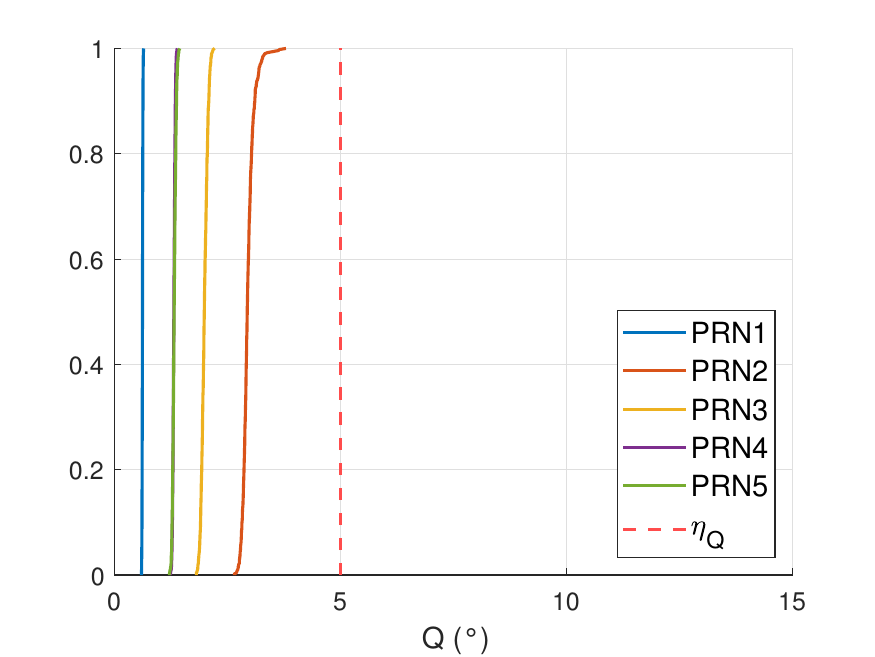}
    \caption{}
    \label{Q_20_cdf}
\end{subfigure}
\caption{CDFs of the $Q$-metric for all satellites: (a) $Q_{0,k}$ at $\theta_0 = 0^\circ$, and (b) $Q_{2,k}$ at $\theta_2 = 20^\circ$.}
\label{cdf}
\end{figure}

For completeness, we illustrate the $Q$-based reliability metric adopted from~\cite{11231075}. 
Fig.~\ref{Q_cdf} shows the cumulative distribution functions (CDFs) of $Q_{0,k}$ at the initial attitude, where Satellites~3 and~5 exhibit noticeably larger $Q$ values than the others. 
Fig.~\ref{Q_20_cdf} shows the CDFs of $Q_{2,k}$ after a $20^{\circ}$ rotation, where all satellites fall within the AMR (see Table~\ref{tab:Visibility_Status}) and yield uniformly low $Q$ values. 
This demonstrates that the $Q$-metric effectively reflects the reliability of the estimates and can guide the reliability-aware selection of estimates for subsequent fusion.

\begin{figure}[h]
\centering
\includegraphics[width=2in]{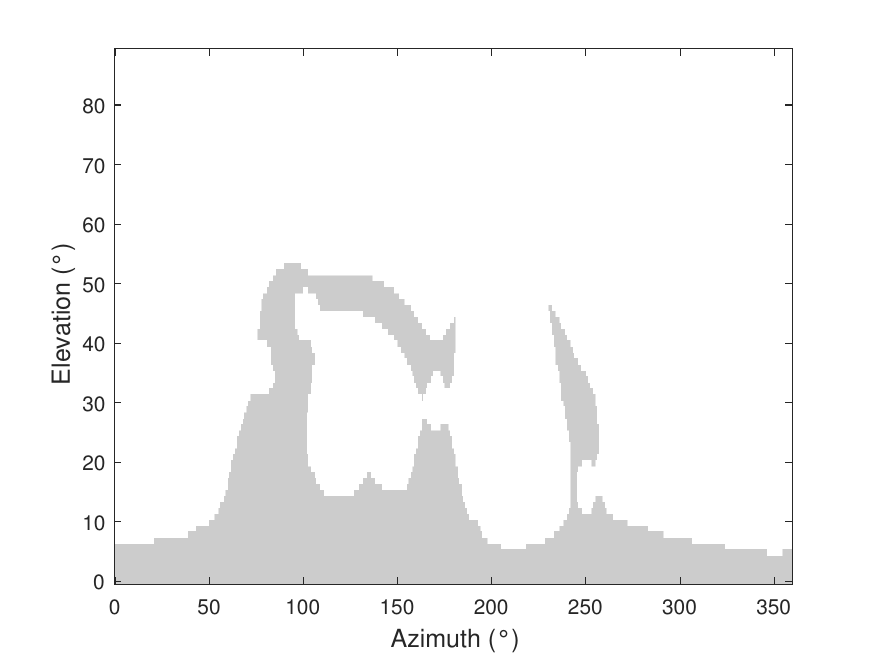}
\caption{Residual IMR formed by the intersection of IMRs across different array attitudes.}
\label{IMR_res}
\end{figure}

It should be noted that the proposed framework does not guarantee accurate estimation for all directions. Fig.~\ref{IMR_res} illustrates the intersection of all IMRs in Fig.~\ref{figQ}, corresponding to directions that remain unresolved under all array attitudes. These directions are typically located near the jamming subspace and are less sensitive to array rotation.

\subsection{Impact of Fusion Strategy and Rotation Increment}

\begin{figure}[h]
\centering
\includegraphics[width=0.8\columnwidth]{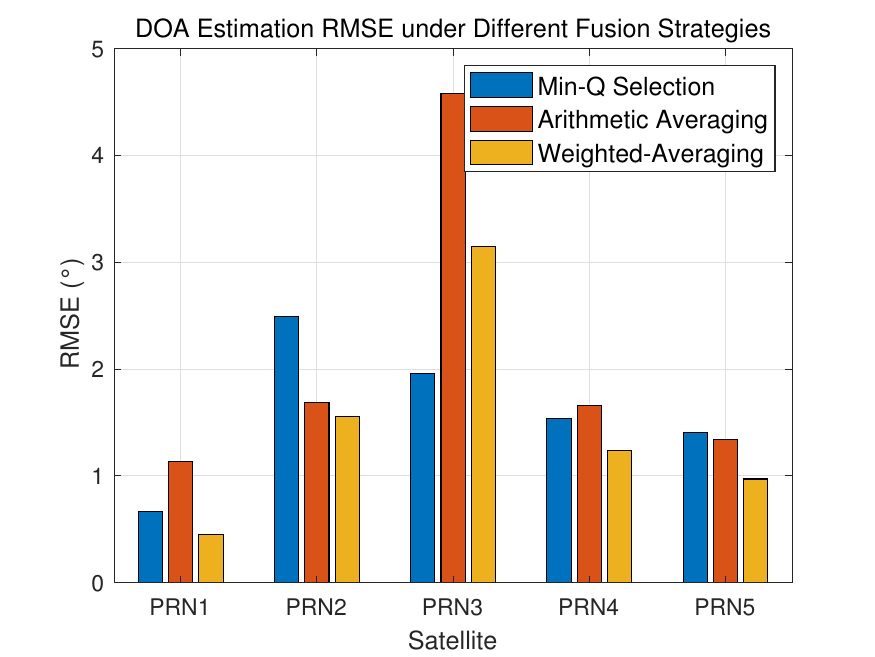}
\caption{Comparison of DOA estimation RMSE for each satellite under three different fusion strategies.}
\label{stratege_compare}
\end{figure}

With the fixed attitude set $\theta_m \in \{0^{\circ}, 10^{\circ}, \dots, 50^{\circ}\}$, we compare the impact of three fusion strategies on DOA estimation accuracy, as shown in Fig.~\ref{stratege_compare}. 
Strategy~3 (weighted averaging) consistently yields lower RMSE than Strategy~2 (arithmetic averaging), highlighting the benefit of assigning higher weights to estimates with lower $Q$ values. 
In contrast, Strategy~1 (Min-$Q$ selection) shows less consistent performance, although it remains competitive for certain satellites, such as Satellite~3. 
Given its low complexity and reasonable accuracy, Strategy~1 remains a practical option in resource-constrained scenarios, while Strategy~3 offers superior robustness across satellites.

\begin{figure}[h]
\centering
\includegraphics[width=0.75\columnwidth]{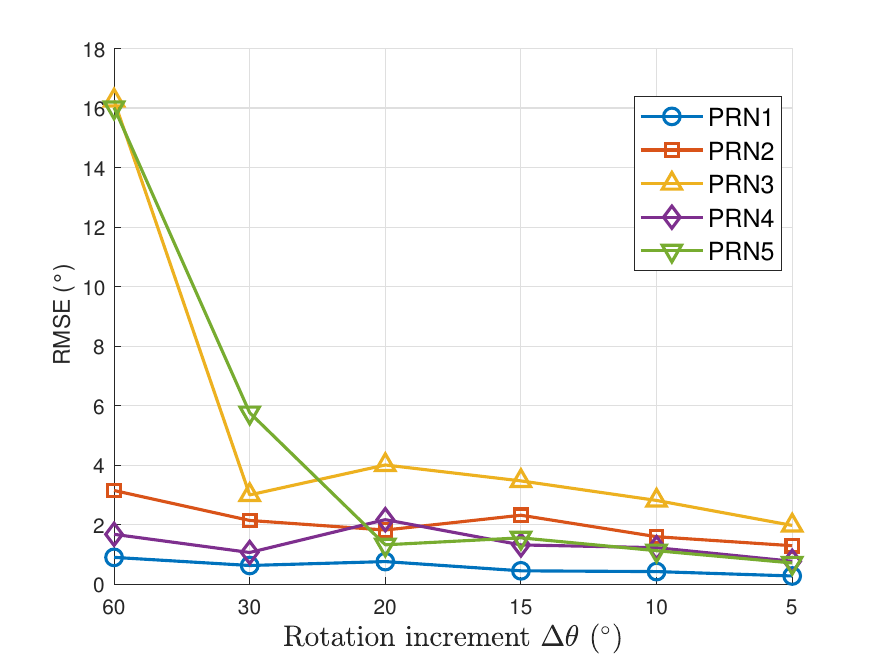}
\caption{Impact of rotation increment on the DOA estimation RMSE for each satellite. }
\label{increment_compare}
\end{figure}

Owing to its superior robustness, Strategy~3 (weighted averaging) is adopted as the default scheme to evaluate the impact of the rotation increment $\Delta\theta = \theta_{m} - \theta_{m-1}$. 
Due to the hexagonal array's six-fold symmetry, the $60^\circ$ interval is uniformly partitioned to test different rotation increments. 
Fig.~\ref{increment_compare} shows the RMSE versus $\Delta\theta$, where $\Delta\theta = 60^\circ$ corresponds to SWN-MUSIC performed at the initial attitude, and smaller increments involve multiple attitudes (e.g., $\Delta\theta = 5^\circ$ corresponds to $M=12$).

Satellites~1, 2, and~4 maintain consistently low RMSE, as they are already accurately resolved at the initial attitude. 
In contrast, the RMSE for Satellites~3 and~5 decreases sharply as $\Delta\theta$ is reduced, stabilizing once the rotation is sufficient (e.g., $20^\circ$–$30^\circ$) to bring them into the AMR. 
Excessively small increments yield diminishing returns. These results indicate that, for the hexagonal array considered here, a rotation step between $10^\circ$ and $20^\circ$ offers a practical balance between estimation accuracy and computational overhead.

\section{Conclusion} \label{sec:conclusion}
We develop a new Rotation-Enhanced SWN-MUSIC framework for robust DOA estimation of GNSS-like signals under jamming, mitigating the effects of the DIP.  
By identifying two types of unfavorable geometric alignments between the array manifold and the jamming subspace, we, for the first time, reveal the underlying mechanisms of the DIP.
The proposed framework exploits the rotational diversity to reconfigure the observation geometry, performing independent estimations across multiple attitudes.  
Simulation results confirm that the approach substantially improves estimation accuracy in directions problematic for fixed-array SWN-MUSIC, without degrading accuracy in previously well-estimated directions.

In summary, our approach effectively expands the performance boundaries of static arrays under challenging jamming conditions, enabling more robust and reliable DOA estimation.
More broadly, it advocates a proactive design principle for array processing: instead of passively accepting the observation geometry of a static array, we can actively manipulate its attitude to ensure accurate estimation results.

\section*{Appendix: Array Attitude Definition and Coordinate Transformations}
\label{Appendix}

\begin{figure}[h]
\centering
\includegraphics[width=0.8\columnwidth]{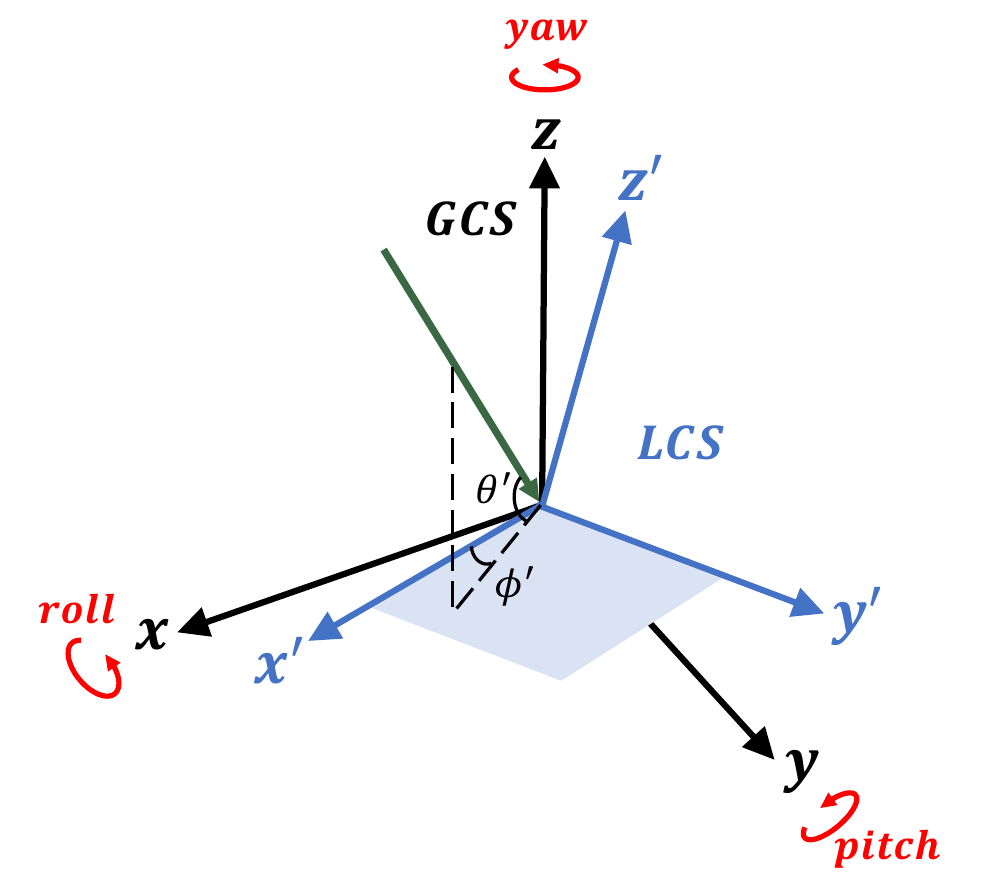}
\caption{Definition of the array attitude using yaw, pitch, and roll Euler angles.}
\label{appendix1}
\end{figure}

As shown in Fig.\ref{appendix1}, the array's attitude is characterized by the Euler angle vector $\boldsymbol{A}=[y, p, r]^{\mathrm{T}}$, representing yaw, pitch, and roll, respectively. 
These angles define a sequential $Z$-$Y$-$X$ rotation from the GCS to the LCS, resulting in the rotation matrix
\begin{equation}
\label{R_matrix}
\boldsymbol{R}(\boldsymbol{A})=\begin{bmatrix} 
c_y c_p & c_y s_p s_r - s_y c_r & c_y s_p c_r + s_y s_r \\ 
s_y c_p & s_y s_p s_r + c_y c_r & s_y s_p c_r - c_y s_r \\ 
-s_p & c_p s_r & c_p c_r 
\end{bmatrix},
\end{equation}
where $c_{\psi}=\cos\psi$ and $s_{\psi}=\sin\psi$ for $\psi \in \{y, p, r\}$. 

Let $\boldsymbol{\gamma} = [\phi, \theta]^\mathrm{T}$ and $\boldsymbol{\gamma}' = [\phi', \theta']^\mathrm{T}$ denote the DOAs of the same signal referenced to the GCS and the current LCS, respectively. 
Their unit direction vectors are related by the following transformation:
\begin{equation}
\boldsymbol{d}(\boldsymbol{\gamma}) = \boldsymbol{R}(\boldsymbol{A}) \boldsymbol{d}(\boldsymbol{\gamma}'),
\end{equation}

\begin{equation}
\boldsymbol{d}(\boldsymbol{\gamma}') = \boldsymbol{R}(\boldsymbol{A})^ \mathrm{T}\boldsymbol{d}(\boldsymbol{\gamma}).
\end{equation}

Consequently, when the array is reconfigured to a known attitude $\boldsymbol{A}$, the DOA estimated in the current LCS can be transformed back into the GCS via:
\begin{equation}
\boldsymbol{\gamma} = \boldsymbol{d}^{-1} \left( \boldsymbol{R}(\boldsymbol{A}) \boldsymbol{d}(\boldsymbol{\gamma}') \right),
\end{equation}
where $\boldsymbol{d}^{-1}(\cdot)$ denotes the inverse mapping from a direction vector to the DOA.

\begin{figure}[h]
\centering
\includegraphics[width=0.8\columnwidth]{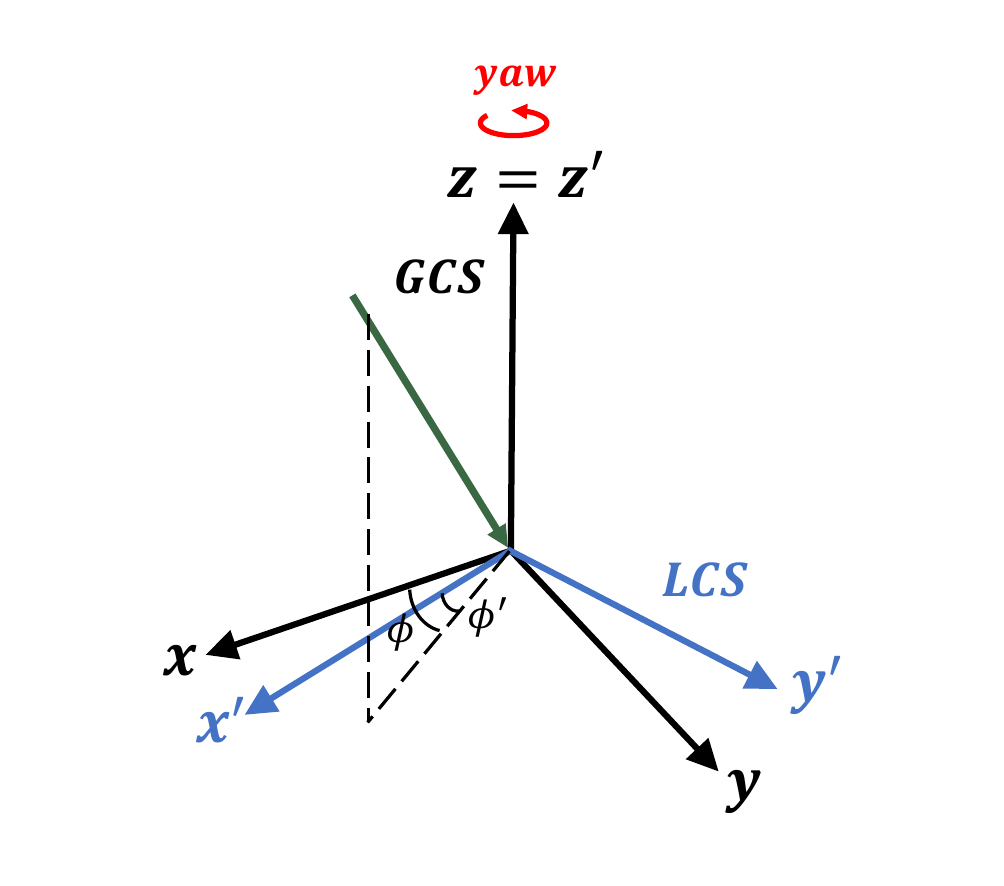}
\caption{Illustration of how the array’s in-plane rotation affects the azimuth in the LCS.  The azimuth in the GCS is obtained by adding the rotation angle to the azimuth estimated in the LCS.}
\label{appendix2}
\end{figure}
In the proposed framework, the attitude adjustment is restricted to in-plane rotation about the array boresight (yaw-only), i.e., $\boldsymbol{A} = [y, 0, 0]^\mathrm{T}$, yielding:
\begin{equation}
\begin{bmatrix} \cos\phi\cos\theta \\ \sin\phi\cos\theta \\ \sin\theta \end{bmatrix} = 
\begin{bmatrix} \cos y & -\sin y & 0 \\ \sin y & \cos y & 0 \\ 0 & 0 & 1 \end{bmatrix}
\begin{bmatrix} \cos\phi'\cos\theta' \\ \sin\phi'\cos\theta' \\ \sin\theta' \end{bmatrix}.
\end{equation}
It is straightforward to obtain:
\begin{equation}
\label{yaw_relation}
\boldsymbol{\gamma} = \boldsymbol{\gamma}' + [y, 0]^\mathrm{T}.
\end{equation}
Therefore, the GCS-referenced DOA is recovered by simply adding the known rotation angle to the estimated azimuth,  as illustrated in Fig.~\ref{appendix2}.

\section*{Acknowledgment}

The authors acknowledge the use of the AI language model ChatGPT (OpenAI) to improve the English language expression throughout the manuscript. No technical content or scientific ideas were generated by the AI.

\bibliographystyle{IEEEtran} 
\bibliography{IEEEabrv,reference}

@ARTICLE{4745647,
  author={Hegarty, Christopher J. and Chatre, Eric},
  journal={Proceedings of the IEEE}, 
  title={Evolution of the Global Navigation SatelliteSystem (GNSS)}, 
  year={2008},
  volume={96},
  number={12},
  pages={1902-1917},
  doi={10.1109/JPROC.2008.2006090}}

@INPROCEEDINGS{11165996,
  author={Noah, Adel and Elmusrati, Mohammed},
  booktitle={2025 International Conference on Artificial Intelligence, Computer, Data Sciences and Applications (ACDSA)}, 
  title={GNSS Spoofing and Jamming Mitigation: A Comprehensive Review}, 
  year={2025},
  volume={},
  number={},
  pages={1-9},
  doi={10.1109/ACDSA65407.2025.11165996}}

@Article{s19183870,
AUTHOR = {Zhang, Jiaqi and Cui, Xiaowei and Xu, Hailong and Lu, Mingquan},
TITLE = {A Two-Stage Interference Suppression Scheme Based on Antenna Array for GNSS Jamming and Spoofing},
JOURNAL = {Sensors},
VOLUME = {19},
YEAR = {2019},
NUMBER = {18},
ARTICLE-NUMBER = {3870},
URL = {https://www.mdpi.com/1424-8220/19/18/3870},
PubMedID = {31500334},
ISSN = {1424-8220},
DOI = {10.3390/s19183870}
}

@ARTICLE{7444116,
  author={Fernández-Prades, Carles and Arribas, Javier and Closas, Pau},
  journal={Proceedings of the IEEE}, 
  title={Robust GNSS Receivers by Array Signal Processing: Theory and Implementation}, 
  year={2016},
  volume={104},
  number={6},
  pages={1207-1220},
  doi={10.1109/JPROC.2016.2532963}}

@article{meurer2016direction,
  title={Direction-of-arrival assisted sequential spoofing detection and mitigation},
  author={Meurer, Michael and Konovaltsev, Andriy and Appel, Manuel and Cuntz, Manuel},
  year={2016}
}

@ARTICLE{10568334,
  author={Wang, Hao and Li, Hong and Zhong, Minghan and Lu, Mingquan},
  journal={IEEE Transactions on Aerospace and Electronic Systems}, 
  title={A Space–Time-Ambiguity Decomposition Method for DOA Estimation Enhancing Antispoofing via Rotating Dual Antennas}, 
  year={2024},
  volume={60},
  number={6},
  pages={7643-7662},
  doi={10.1109/TAES.2024.3417434}}

@inproceedings{meurer2012robust,
  title={Robust joint multi-antenna spoofing detection and attitude estimation using direction assisted multiple hypotheses RAIM},
  author={Meurer, Michael and Konovaltsev, Andriy and Cuntz, Manuel and H{\"a}ttich, Christian},
  booktitle={proceedings of the 25th international technical meeting of the satellite division of the Institute of Navigation (ION GNSS 2012)},
  pages={3007--3016},
  year={2012}
}

@ARTICLE{11231075,
  author={Wang, Chuanrui and Cui, Xiaowei and Liu, Gang and Lu, Mingquan},
  journal={IEEE Transactions on Aerospace and Electronic Systems}, 
  title={Attitude Determination Using GNSS DOA Estimation Under Jamming}, 
  year={2026},
  volume={62},
  number={},
  pages={1464-1477},
  doi={10.1109/TAES.2025.3630146}}

@book{markel2002interference,
  title={Interference mitigation for GPS based attitude determination},
  author={Markel, Matthew David},
  year={2002},
  publisher={University of Florida}
}

@inproceedings{arribas2023receiver,
  title={A receiver-independent GNSS smart antenna for simultaneous jamming and spoofing protection},
  author={Arribas, Javier and G{\'o}mez, MA and Fern{\'a}ndez-Prades, Carles and Mart{\'\i}n, David Laso and Garc{\'\i}a-Tu{\~n}{\'o}n, Julio Mar{\'\i}a and Rioja, Tamara Garc{\'\i}a},
  booktitle={2023 IEEE Aerospace Conference},
  pages={1--13},
  year={2023},
  organization={IEEE}
}

@article{wang2024two,
  title={Two-stage spatial whitening and normalized MUSIC for robust DOA estimation of GNSS signals under jamming},
  author={Wang, Chuanrui and Cui, Xiaowei and Liu, Gang and Lu, Mingquan},
  journal={IEEE Transactions on Aerospace and Electronic Systems},
  year={2024},
  publisher={IEEE}
}

@article{sun2022anti,
  title={Anti-jamming method and implementation for GNSS receiver based on array antenna rotation},
  author={Sun, Yifan and Chen, Feiqiang and Lu, Zukun and Wang, Feixue},
  journal={Remote Sensing},
  volume={14},
  number={19},
  pages={4774},
  year={2022},
  publisher={MDPI}
}

@electronic{AntcomCRPA,
  author       = {{Antcom Corporation}},
  title        = {Controlled Reception Pattern Antennas ({CRPAs})},
  url          = {https://www.antcom.com/products/crpas},
  note         = {[Accessed: 06-Feb-2026]},
  year         = {2026}
}

@electronic{HarxonAntenna,
  author       = {{Harxon Corporation}},
  title        = {{GNSS} Antennas},
  url          = {https://www.harxon.com/product/63},
  note         = {[Accessed: 06-Feb-2026]},
  year         = {2026}
}

@article{akos2012s,
  title={Who's afraid of the spoofer? GPS/GNSS spoofing detection via automatic gain control (AGC)},
  author={Akos, Dennis M},
  journal={NAVIGATION: Journal of the Institute of Navigation},
  volume={59},
  number={4},
  pages={281--290},
  year={2012},
  publisher={Wiley Online Library}
}

@inproceedings{deng2022gnss,
  title={GNSS spoofing detection based on abnormal receiver noise and carrier-to-noise ratio metric},
  author={Deng, Min and Wang, Honghao and Ming, Dexiang and Chen, Yangzhuo},
  booktitle={2022 IEEE 22nd International Conference on Communication Technology (ICCT)},
  pages={1306--1311},
  year={2022},
  organization={IEEE}
}

@article{zhou2022new,
  title={A new induced GNSS spoofing detection method based on weighted second-order central moment},
  author={Zhou, Wenlong and Lv, Zhiwei and Deng, Xu and Ke, Ye},
  journal={IEEE sensors journal},
  volume={22},
  number={12},
  pages={12064--12078},
  year={2022},
  publisher={IEEE}
}

@article{chen2022gnss,
  title={GNSS multiparameter spoofing detection method based on support vector machine},
  author={Chen, Zhengkun and Li, Junzhi and Li, Jing and Zhu, Xiangwei and Li, Caihua},
  journal={IEEE Sensors Journal},
  volume={22},
  number={18},
  pages={17864--17874},
  year={2022},
  publisher={IEEE}
}

@article{broumandan2016overview,
  title={Overview of spatial processing approaches for GNSS structural interference detection and mitigation},
  author={Broumandan, Ali and Jafarnia-Jahromi, Ali and Daneshmand, Saeed and Lachapelle, G{\'e}rard},
  journal={Proceedings of the IEEE},
  volume={104},
  number={6},
  pages={1246--1257},
  year={2016},
  publisher={IEEE}
}

@article{jones2017spoofing,
  title={Spoofing in the Black Sea: What really happened?},
  author={Jones, Michael},
  journal={GPS World},
  volume={11},
  year={2017}
}

@article{androjna2021impact,
  title={Impact of spoofing of navigation systems on maritime situational awareness},
  author={Androjna, Andrej and Perkovi{\v{c}}, Marko},
  journal={Transactions on Maritime Science},
  volume={10},
  number={02},
  pages={361--373},
  year={2021},
  publisher={Sveu{\v{c}}ili{\v{s}}te u Splitu, Pomorski fakultet}
}

@article{schmidt2016survey,
  title={A survey and analysis of the GNSS spoofing threat and countermeasures},
  author={Schmidt, Desmond and Radke, Kenneth and Camtepe, Seyit and Foo, Ernest and Ren, Micha{\l}},
  journal={ACM Computing Surveys (CSUR)},
  volume={48},
  number={4},
  pages={1--31},
  year={2016},
  publisher={ACM New York, NY, USA}
}

@inproceedings{appel2015robust,
  title={Robust spoofing detection and mitigation based on direction of arrival estimation},
  author={Appel, Manuel and Konovaltsev, Andriy and Meurer, Michael},
  booktitle={Proceedings of the 28th International Technical Meeting of The Satellite Division of the Institute of Navigation (ION GNSS+ 2015)},
  pages={3335--3344},
  year={2015}
}

@ARTICLE{10480364,
  author={Liu, Jinyuan and Chen, Feiqiang and Xie, Yuchen and Ge, Beibei and Lu, Zukun and Sun, Guangfu},
  journal={IEEE Sensors Journal}, 
  title={Robust Spoofing Detection for GNSS Array Receivers Using the Auxiliary Reference Element Method}, 
  year={2024},
  volume={24},
  number={9},
  pages={14833-14843},
  doi={10.1109/JSEN.2024.3379848}}

@ARTICLE{11159122,
  author={Wang, Hao and Chang, Qing and Liu, Yongqiang and Zhao, Hongbo and Zhuang, Chen},
  journal={IEEE Transactions on Vehicular Technology}, 
  title={A GNSS Anti-Spoofing Method Based on Parallel Tracking Processing Under Array Antenna Architecture}, 
  year={2026},
  volume={75},
  number={3},
  pages={4760-4770},
  doi={10.1109/TVT.2025.3608897}}

@inproceedings{danelli2026gnss,
  author    = {Danelli, L. and Emmanuele, A. and Pastori, N. and Colombo, R. and Iacone, P. and Marradi, L. and Siniscalco, L.},
  title     = {{GNSS} Spoofing Detection and {DOA} estimation with an Array Receiver and its Digital Twin: Field Test at {Jammertest} 2025 in {Norway}},
  booktitle = {Proceedings of the 2026 International Technical Meeting of The Institute of Navigation},
  year      = {2026},
  address   = {Anaheim, California},
  month     = {January},
  pages     = {344--358},
  doi       = {10.33012/2026.20543},
  url       = {https://doi.org/10.33012/2026.20543}
}

@ARTICLE{7444122,
  author={Amin, Moeness G. and Wang, Xiangrong and Zhang, Yimin D. and Ahmad, Fauzia and Aboutanios, Elias},
  journal={Proceedings of the IEEE}, 
  title={Sparse Arrays and Sampling for Interference Mitigation and DOA Estimation in GNSS}, 
  year={2016},
  volume={104},
  number={6},
  pages={1302-1317},
  doi={10.1109/JPROC.2016.2531582}}

@Article{rs15235497,
AUTHOR = {Liu, Rui and Yang, Zhiwei and Chen, Qidong and Liao, Guisheng and Zhu, Qinglin},
TITLE = {Localization of GNSS Spoofing Interference Source Based on a Moving Array Antenna},
JOURNAL = {Remote Sensing},
VOLUME = {15},
YEAR = {2023},
NUMBER = {23},
ARTICLE-NUMBER = {5497},
URL = {https://www.mdpi.com/2072-4292/15/23/5497},
ISSN = {2072-4292},
DOI = {10.3390/rs15235497}
}

@inproceedings{daneshmand2012low,
  title={A low-complexity GPS anti-spoofing method using a multi-antenna array},
  author={Daneshmand, Saeed and Jafarnia-Jahromi, Ali and Broumandon, Ali and Lachapelle, G{\'e}rard},
  booktitle={Proceedings of the 25th international technical meeting of the satellite division of the institute of navigation (ION GNSS 2012)},
  pages={1233--1243},
  year={2012}
}

@inproceedings{Danelli2026,
  author    = {Danelli, L. and Emmanuele, A. and Pastori, N. and Colombo, R. and Iacone, P. and Marradi, L. and Siniscalco, L.},
  title     = {GNSS Spoofing Detection and DOA Estimation with an Array Receiver and its Digital Twin: Field Test at Jammertest 2025 in Norway},
  booktitle = {Proceedings of the 2026 International Technical Meeting of The Institute of Navigation},
  year      = {2026},
  pages     = {344--358},
  address   = {Anaheim, California},
  month     = {January},
  doi       = {10.33012/2026.20543},
  url       = {https://doi.org/10.33012/2026.20543}
}

@inproceedings{marcos2018interference,
  title={Interference and spoofing detection for GNSS maritime applications using direction of arrival and conformal antenna array},
  author={Marcos, E P{\'e}rez and Konovaltsev, Andriy and Caizzone, Stefano and Cuntz, Manuel and Yinusa, Kazeem and Elmarissi, Wahid and Meurer, Michael},
  booktitle={Proceedings of the 31st International Technical Meeting of The Satellite Division of the Institute of Navigation (ION GNSS+ 2018)},
  pages={2907--2922},
  year={2018}
}

@ARTICLE{7445815,
  author={Psiaki, Mark L. and Humphreys, Todd E.},
  journal={Proceedings of the IEEE}, 
  title={GNSS Spoofing and Detection}, 
  year={2016},
  volume={104},
  number={6},
  pages={1258-1270},
  doi={10.1109/JPROC.2016.2526658}}

@article{radovs2024recent,
  title={Recent advances on jamming and spoofing detection in GNSS},
  author={Rado{\v{s}}, Katarina and Brki{\'c}, Marta and Begu{\v{s}}i{\'c}, Dinko},
  journal={Sensors},
  volume={24},
  number={13},
  pages={4210},
  year={2024},
  publisher={MDPI}
}

@inproceedings{Lo2026Global,
  author    = {Lo, Sherman and Liu, Zixi and Ibrahim, Lyla and Chen, Yu Hsuan and Akos, Dennis and Walter, Todd},
  title     = {Global Incidents of Aviation Spoofing in 2024-2025 Detected with Automatic Dependent Surveillance Broadcast},
  booktitle = {Proceedings of the 2026 International Technical Meeting of the Institute of Navigation},
  address   = {Anaheim, California},
  month     = jan,
  year      = {2026},
  pages     = {502--515},
  doi       = {10.33012/2026.20527},
}

@ARTICLE{10.3389/fphy.2023.1133316,
    
AUTHOR={Qiao, Jia  and Lu, Zukun  and Lin, Baojun  and Song, Jie  and Xiao, Zhibin  and Wang, Zhi  and Li, Baiyu },
           
TITLE={A survey of GNSS interference monitoring technologies},
          
JOURNAL={Frontiers in Physics},
          
VOLUME={Volume 11 - 2023},
  
YEAR={2023},
  
URL={https://www.frontiersin.org/journals/physics/articles/10.3389/fphy.2023.1133316},
  
DOI={10.3389/fphy.2023.1133316},
  
ISSN={2296-424X}}

@ARTICLE{7444136,
  author={Gao, Grace Xingxin and Sgammini, Matteo and Lu, Mingquan and Kubo, Nobuaki},
  journal={Proceedings of the IEEE}, 
  title={Protecting GNSS Receivers From Jamming and Interference}, 
  year={2016},
  volume={104},
  number={6},
  pages={1327-1338},
  doi={10.1109/JPROC.2016.2525938}}

@ARTICLE{10719606,
  author={Li, Song and Lin, Honglei and Tang, Xiaomei and Wang, Feixue},
  journal={IEEE Sensors Journal}, 
  title={Blind Spoofing Detection for Anti-Jamming Multiantenna GNSS Receivers}, 
  year={2024},
  volume={24},
  number={23},
  pages={39418-39431},
  doi={10.1109/JSEN.2024.3477309}}

\begin{IEEEbiography}[{\includegraphics[width=1in,height=1.25in,clip,keepaspectratio]{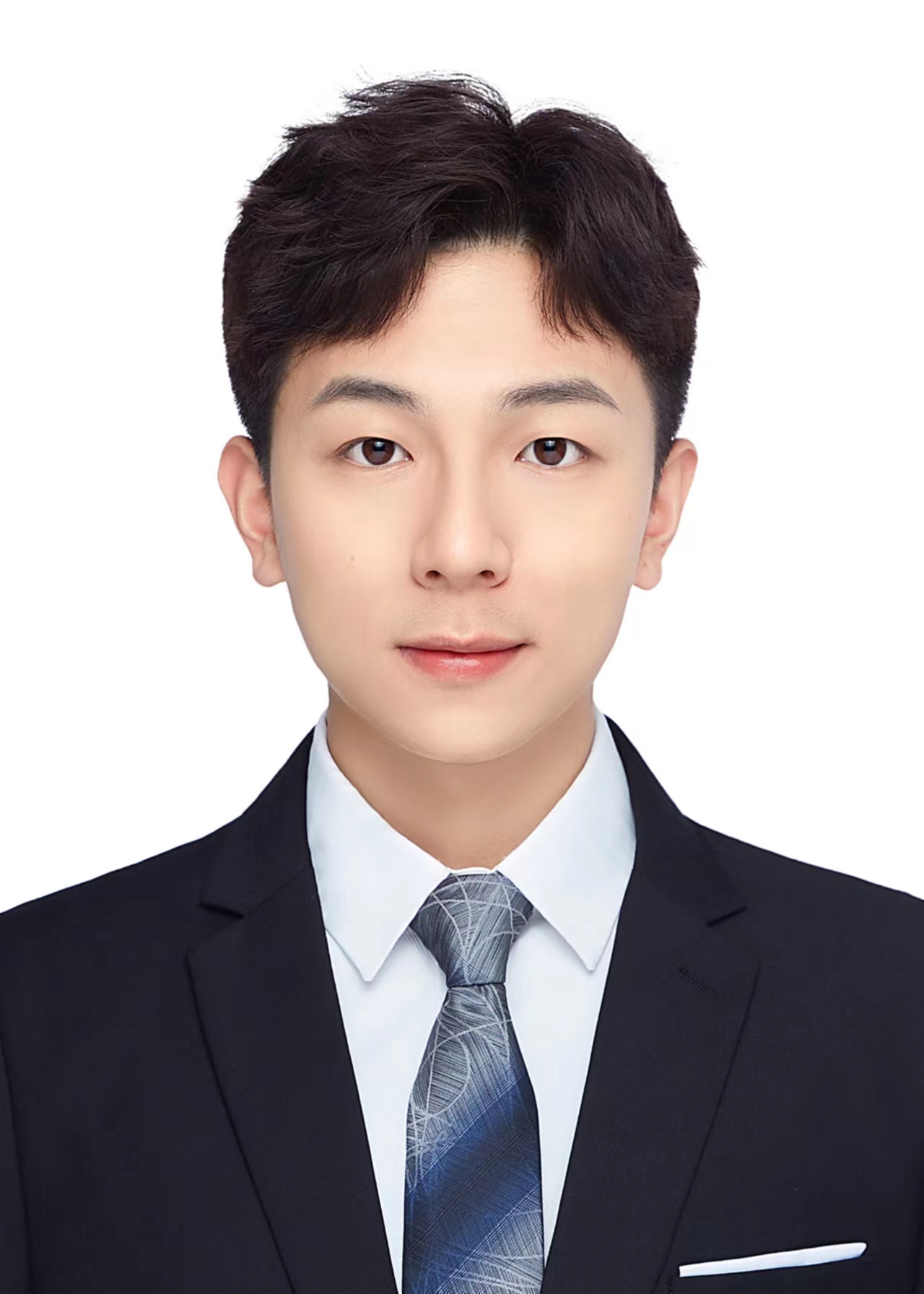}}]{Jiyao Jiang}
is a Ph.D. candidate at the Department of Electronic Engineering at Tsinghua University, China. He received his B.E. degree in Electronic Engineering from Tsinghua University, Beijing, China, in 2023. His current research focuses on GNSS array signal processing.
\end{IEEEbiography}

% \begin{IEEEbiography}[{\includegraphics[width=1in,height=1.25in,clip,keepaspectratio]{Wang}}]{Chuanrui Wang}
% is a Ph.D. candidate at the Department of Electronic Engineering at Tsinghua University, China. He received his B.E. degree in Electronic Engineering from Tsinghua University, Beijing, China, in 2021. His current research focuses on GNSS array signal processing.
% \end{IEEEbiography}

\begin{IEEEbiography}[{\includegraphics[width=1in,height=1.25in,clip,keepaspectratio]{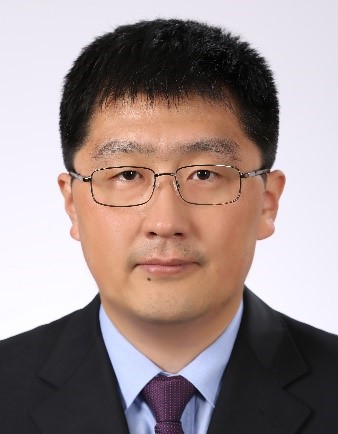}}]{Xiaowei Cui}
received the B.S. and Ph.D. degrees in electronic engineering from Tsinghua
University, Beijing, China, in 2000 and 2005,
respectively. Since 2005, he has been with
the Department of Electronic Engineering, Tsinghua University, where he is currently a Professor.
His research interests include robust GNSS
signal processing, multipath mitigation techniques, and high-precision positioning. He is
a member of the Expert Group of China BeiDou Navigation Satellite
System.
\end{IEEEbiography}

\begin{IEEEbiography}[{\includegraphics[width=1in,height=1.25in,clip,keepaspectratio]{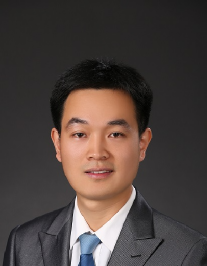}}]{Gang Liu} 
is an associate researcher at the Department of Electronic Engineering, Tsinghua University, China. His research interests include GNSS/INS integrated navigation techniques, and high-precision localization. He obtained both the BS and PhD degrees in instrument science and technology from Tsinghua University in 2007 and 2015, respectively.
\end{IEEEbiography}

\begin{IEEEbiography}[{\includegraphics[width=1in,height=1.25in,clip,keepaspectratio]{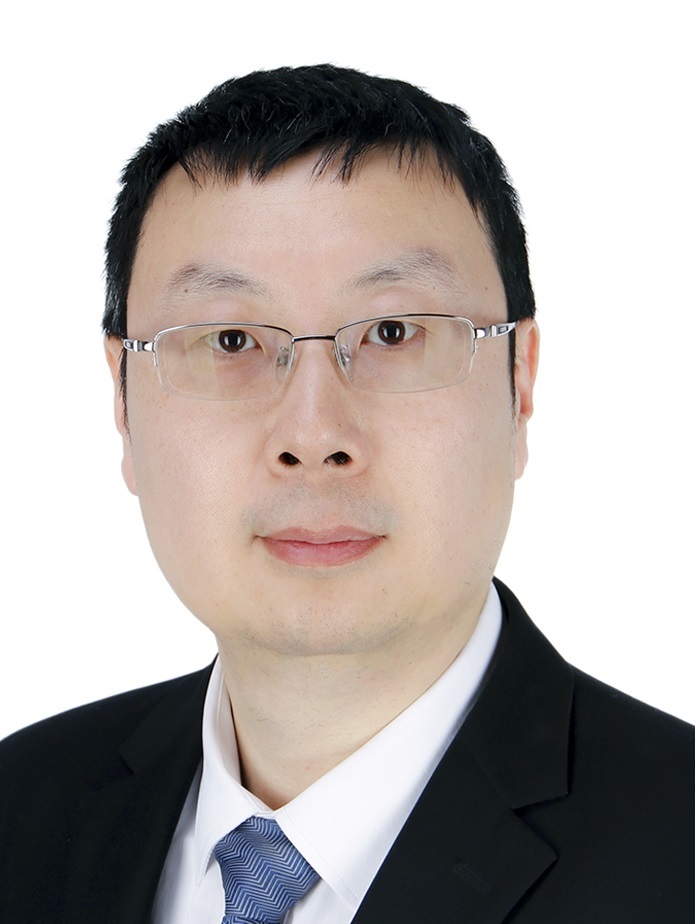}}]{Sihao Zhao} 
 (Senior Member, IEEE)
received the B.S. and Ph.D. degrees in electronic engineering from Tsinghua University, Beijing, China, in 2005 and 2011, respectively. He is currently a Senior Staff Engineer with etherWhere Corporation, CA, USA. From 2011 to 2013, he was an Electronics Systems Engineer with China Academy of Space Technology, Beijing, China. From 2013 to 2019, he was a Postdoctoral Researcher and then an Assistant Professor with the Department of Electronic Engineering, Tsinghua University, Beijing, China. From 2020 to 2022, he was a Research Associate with the Department of Electrical, Computer and Biomedical Engineering, Toronto Metropolitan University, Toronto, ON, Canada. From 2022 to 2024, he was a Senior Algorithm Designer with NovAtel, Autonomy and Positioning division of Hexagon, AB, Canada. He is an Editor of GPS Solutions.
\end{IEEEbiography}

\begin{IEEEbiography}[{\includegraphics[width=1in,height=1.25in,clip,keepaspectratio]{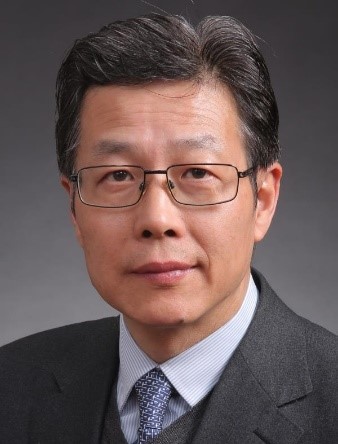}}]{Mingquan Lu}
received the M.S. and Ph.D.
degrees from University of Electronic Science
and Technology of China, Chengdu, China.
He is a Professor with the Department of
Electronic Engineering, Tsinghua University,
Beijing, China. He directs the PNT Research
Center, which develops GNSS and alternative
PNT technologies. He authored or co-authored
5 books and book chapters, published over 300
journal and conference papers, and hold nearly
100 patents. His current research focuses on GNSS signal processing,
GNSS receiver development and emerging PNT technologies.
Prof. Lu was a recipient of the ION Thurlow Award. He provided
numerous services to the GNSS community, including an Associate Editor for Satellite Navigation and Journal of Navigation and Positioning.
He is a Fellow of ION.
\end{IEEEbiography}

\end{document}